\documentclass[a4paper]{article}
\usepackage[a4paper]{geometry}

\usepackage{amssymb,amsmath,multirow,booktabs,float,graphicx}
\usepackage{cite}

\usepackage[format=hang]{caption}
\numberwithin{equation}{section}

\title{Computation of Dynamical Correlation Functions of the Spin-1 Babujan-Takhtajan Chain}
\author{R.P.Vlijm and J.-S. Caux \\ \small Institute for Theoretical Physics, University of Amsterdam,\\ \small Science Park 904, Postbus 94485, 1090 GL Amsterdam, The Netherlands\\ \small E-mail: r.p.vlijm@uva.nl and j.s.caux@uva.nl}
\begin{document}
\maketitle
\begin{abstract}The dynamical structure factor of the Babujan-Takhtajan antiferromagnetic spin-1 chain is computed numerically at zero temperature and zero magnetic field, using the higher-spin generalization of an Algebraic Bethe Ansatz-based method previously used for spin-$1/2$ integrable chains. This method, which consists in the explicit construction of eigenstates and the summation of the Lehmann representation of the correlator, is here particularly challenging to implement in view of the presence of strongly deviated string solutions to the Bethe equations. We show that a careful treatment of these deviations makes it possible to obtain perfect saturation of sum rules for small system sizes, and extremely good saturation for large system sizes where the dynamical structure factor is computed by including all two-spinon and four-spinon contributions. The real-space spin-spin correlation, obtained by Fourier transforming our results, displays asymptotics fitting predictions from conformal field theory.
\end{abstract}

\section{Introduction}
Quantum spin chains in one dimension are of interest both from theoretical and experimental perspectives. The prototypical spin-$\frac{1}{2}$ Heisenberg model is integrable and its eigenstates can be obtained exactly using Bethe Ansatz \cite{1931_Bethe_ZP_71}. Moreover, the development of the Algebraic Bethe Ansatz (see \cite{KorepinBOOK}) gave access to the computation of experimentally relevant quantities such as dynamical spin-spin correlation functions. In particular, the existence of a determinant representation of the scalar product between Bethe states \cite{1989_Slavnov_TMP_79,1990_Slavnov_TMP_82} and of matrix elements of local operators \cite{1999_Kitanine_NPB_554} opens the way towards an explicit summation of spectral representations for such correlators. Recent inelastic neutron scattering experiments for the antiferromagnetic spin-$\frac{1}{2}$ Heisenberg chain~\cite{2013_Mourigal_NATPHYS_9,2013_Lake_PRL_111} have displayed very good comparison with theoretical predictions following from Algebraic Bethe Ansatz on both the spinon structure of the low-lying excitations and dynamical correlations. 

The unifying view of integrable models afforded by the Algebraic Bethe Ansatz means that such methods can be extended to other models. In the context of spin chains, a generalisation of the Heisenberg model to integrable chains of higher spin is possible through the fusion of R-matrices~\cite{1981_Kulish_LMP_5}. The aim of this paper is to extend the computation of dynamical correlation functions for the Heisenberg model to the integrable spin-$1$ case, namely to the Babujan-Takhtajan \cite{1982_Takhtajan_PLA_87,1982_Babujan_PLA_90,1983_Babujan_NPB_215} bilinear-biquadratic model (see equation~(\ref{eq:HamiltonianBT}) below). 

A straightforward generalisation of the Heisenberg ({\it i.e.} strictly bilinear) model to higher spin leads to a model with finite Haldane gap~\cite{1983_Haldane_PRL_50,1983_Haldane_PLA_93} for integer spin, while it remains gapless for half-integer spin. In both cases the Heisenberg model is not integrable away from spin-$\frac{1}{2}$, and the simplest higher-spin extensions consider a Hamiltonian which is polynomial in the nearest-neighbour interaction terms. For the first possible generalisation of the Heisenberg chain to spin-$1$, a biquadratic nearest-neighbour interaction term in the Hamiltonian can be added, which for different ratios between the coefficients of the bilinear and biquadratic terms leads to a rich phase diagram full of interesting physical properties~\cite{1986_Affleck_NPB_265,1987_Solyom_PRB_36}. The general bilinear-biquadratic spin-$1$ chain Hamiltonian can be written as
\begin{equation}
H=\frac{J}{4} \sum_j \left[  \hat  {\boldsymbol S}_{j} \cdot \hat {\boldsymbol S}_{j+1} + \alpha (\hat {\boldsymbol S}_{j} \cdot \hat {\boldsymbol S}_{j+1})^{2} \right].
\label{eq:Hamiltonianalpha}
\end{equation}
A complete overview of the phase diagram of the bilinear-biquadratic spin-$1$ Hamiltonian, including the presence of a magnetic field, is given in~\cite{2011_Manmana_PRB_83}. Only a few exactly solvable points are identifiable. Regarding the notion of exact solvability, a point worth mentioning in the phase diagram at $\alpha=\frac{1}{3}$ is known as the AKLT-model~\cite{1987_Affleck_PRL_59,1988_Affleck_CMP_115}. The ground state of this model is exactly expressed as a valence bond solid. However, this model is not Bethe Ansatz-solvable, so its exact spectrum and correlations are out of reach.

The aforementioned R-matrix fusion procedure within the framework of the Algebraic Bethe Ansatz yields the SU($2$)-symmetric integrable higher spin chains. For the specific case of a spin-$1$ chain, the corresponding Hamiltonian is the Babujan-Takhtajan model with $\alpha=-1$, namely
\begin{equation}
H=\frac{J}{4} \sum_j \left[  \hat  {\boldsymbol S}_{j} \cdot \hat {\boldsymbol S}_{j+1} - (\hat {\boldsymbol S}_{j} \cdot \hat {\boldsymbol S}_{j+1})^{2} \right].
\label{eq:HamiltonianBT}
\end{equation}
It is worth mentioning that, by means of the nested Bethe Ansatz, the eigenstates of another integrable spin-$1$ model with $\alpha = 1$ can be constructed. This SU(3)-symmetric chain is known as the Uimin-Lai-Sutherland model\cite{1970_Uimin_JETPL_12,1974_Lai_JMP_15,1975_Sutherland_PRB_12}. There have been recent developments on the computation of scalar products between Bethe states for SU(3) models~\cite{2012_Belliard_JSTAT_P10017}, but a full determinant representation similar to the one existing in the SU(2) case has not yet been found. The procedure of the computation of dynamical correlation functions based on the Algebraic Bethe Ansatz can thus only be extended to the spin-$1$ chain for the Babujan-Takhtajan model at this point.

The Bethe equations for higher integrable spin-$s$ chains have been derived by the fusion of R-matrices. However, the derivation might as well be performed using the Coordinate Bethe Ansatz~\cite{2011_Crampe_SIGMA_7}, following the original derivation of the Bethe equations~\cite{1931_Bethe_ZP_71} for the Heisenberg model by imposing plane wave solutions and periodic boundary conditions. The Bethe equations for the spin-$s$ case are given by
\begin{equation}
\left(\frac{\lambda_{j}+i s}{\lambda_{j}-i s} \right)^{N}=\prod_{k\neq j}^{M}\frac{\lambda_{j}-\lambda_{k}+i}{\lambda_{j}-\lambda_{k}-i}
\label{eq:BEspinS}
\end{equation}
where each set of roots $\lambda_{j}\in\{\lambda_{1},...,\lambda_{M} \}$ specifies an eigenstate of the model. The length of the spin chain is denoted by $N$, while $M$ denotes the number of downturned spins as compared to the fully polarised reference state $|0\rangle=\bigotimes_{j}^{N}|\uparrow\rangle_{j}$. Moreover, the energy of a Bethe state is given by
\begin{equation}
E=-J\sum_{j=1}^{M}\frac{s}{s^2+\lambda_{j}^{2}}.
\end{equation}

The Hamiltonians corresponding to the Bethe equations for a spin-$s$ chain are polynomials of degree $2s$ in the nearest neighbour interaction between spins~\cite{1983_Babujan_NPB_215}. The Babujan-Takhtajan model (\ref{eq:HamiltonianBT}) is thus the first step above the Heisenberg model in this hierarchy.

While the eigenstates of the model are in principle obtained by means of the Bethe Ansatz, solving for the roots of the Bethe equations for each state is still a highly non-trivial task. Expressions for correlation functions of integrable higher spin chains for the isotropic as well as the anisotropic chain at both zero and finite temperature have been obtained by means of the Algebraic Bethe Ansatz, fusion and/or multiple integral representations \cite{2001_Kitanine_JPA_34,2010_Deguchi_NPB_831,2010_Goehmann_JSTAT_P11011}, avoiding the necessity of having to deal with the roots of the Bethe equations explicitly. However, these approaches cannot be applied to the derivation of dynamical correlation functions, which are of great importance from a physical point of view and which we will concentrate on here.

Some properties of the solutions of the Bethe equations might be considered in advance. If the set of Bethe roots $\{\lambda\}$ solves the Bethe equations, then it is straightforward to see that the set $\{\lambda\}^*$ must also be a valid solution by considering the complex conjugated version of the Bethe equations. Moreover, it is possible to prove a stronger statement~\cite{1986_Vladimirov_TMP_66} that the set of roots solving the Bethe equations have the feature of being self conjugate, $\{\lambda\}=\{\lambda\}^*$. This important property boils down to the fact that the Bethe roots organise themselves symmetrically around the real axis of the complex rapidity plane. In general for the thermodynamic limit, the solutions are furthermore conjectured to group themselves in structures called strings, sharing a common real part, but separated in the imaginary direction by $i$. String solutions of length larger then one are then to be interpreted as bound states of local spin lowering operators acting on the reference state of the spin chain. For finite size, the string hypothesis technically fails and one encounters deviations from the assumed perfect string patterns. The only constraint on the solutions is its self conjugacy, yielding a picture of deformed strings as a result of the finite size. The deviations of the string solutions regularise the divergent prefactors and determinants arising in the matrix element expressions.

 For the spin-$\frac{1}{2}$ Heisenberg model, the antiferromagnetic ground state consists of real rapidities only, while the low-lying excitations may contain a small number of strings. Moreover, deviations of single two-strings are in general (at least at finite field) exponentially suppressed in system size, which allows for not taking deviations into account in the numerical solution procedure of the Bethe equations and computation of matrix elements. The expressions for the matrix elements can be rewritten for string solutions by rearranging the determinants and cancelling the divergent terms among each other. For the majority of purposes and in most cases it is therefore sufficient for high precision computations to neglect string deviations in the spin-$\frac{1}{2}$ model.

In contrast to the spin-$\frac{1}{2}$ antiferromagnetic ground state containing only real rapidities, the ground state of the spin-$1$ Babujan-Takhtajan chain is conjectured to consist solely out of two-strings~\cite{1982_Takhtajan_PLA_87,1983_Babujan_NPB_215}. The bilinear term of the Hamiltonian favours antiparallel ordering of neighbouring spins, while the biquadratic term lowers the energy when neighbouring spins are ordered either parallel or antiparallel. Altogether this will result in antiparallel ordering in the ground state, which can be approximated by acting twice with a spin lowering operator on every other site. This locally bound state of two down spins gives a good intuition as to why the ground state Bethe Ansatz solution is a sea of two-strings. The low-lying excited states can subsequently be realised by breaking up one or more two-strings into real rapidties or higher strings.

Most importantly, a thermodynamic number of two-strings is present in the spin-$1$ ground state, and the corresponding string deviations are not exponentially vanishing with respect to system size~\cite{1990_deVega_JPA_23,1990_Kluemper_JPA_23}. A proper description of the string structure in the Bethe states as well as the behaviour of string deviations are an essential part of our computation, due to the presence of strings in both the ground state and the low-lying excited states.
 
The main goal of this paper is to present a computation of the dynamical structure factor of the Babujan-Takhtajan spin-$1$ chain, following the numerical strategy introduced for the spin-$\frac{1}{2}$ Heisenberg model~\cite{2005_Caux_PRL_95,2005_Caux_JSTAT_P09003}, while taking the more involved difficulties of handling string deviations for the spin-$1$ case into account. The dynamical structure factor is defined as the Fourier transform of the connected spin-spin correlation function
\begin{align}
S^{a\bar a}(q,\omega)&=\frac{1}{N}\sum_{j,j^\prime}^N e^{i q (j-j^\prime)}\int_{-\infty}^\infty \text{d}t \; e^{i\omega t} \langle S_j^a (t) S_{j^\prime}^{\bar a} (0)\rangle_c \label{eq:defDSF1}\\
&=2\pi \sum_\alpha | \langle \text{GS} | S^a_q | \alpha \rangle |^2 \delta(\omega-\omega_\alpha). \label{eq:defDSF2}
\end{align}
From the first to the second line we switch to the Lehmann representation by making use of a resolution of the identity $1=\sum_\alpha | \alpha \rangle \langle \alpha | $ to obtain a sum over intermediate states. For the zero temperature dynamical structure factor, the expectation value with respect to the ground state should be taken. Furthermore, as we take the connected correlator, the ground state is excluded from the sum over intermediate states.
One can distinguish between the dynamical structure factor in the transverse ($a=\pm$) and longitudinal ($a=z$) direction of the spin chain, although this distinction is immaterial in the zero-field isotropic case we consider here.
Besides a careful analysis and proper calculation of the deviated string solutions, the computation of the dynamical structure factor relies on determinant representations for matrix element expressions obtained for higher spin chains~\cite{2007_Castro_Alvaredo_JPA_40}.

The paper is structured in the following way. The next section will provide an overview of the structure of the Bethe Ansatz solutions for the Babujan-Tahktajan spin-$1$ chain, while the third section will introduce the parametrisation and method to obtain the roots of the Bethe equations including string deviations. Section 4 will elaborate on the matrix element expressions for spin-$1$, while the results will be given in section 5. 

\section{Structure of solutions}
\label{sec:two}
In order to investigate the structure of the eigenstates of the higher spin chains, it is more convenient to introduce the logarithm of the Bethe equations. The logarithmic branches will provide for quantum numbers which specify the states. The logarithmic Bethe equations for the integrable spin-$s$ chain are given by
\begin{equation}
\theta_{2s}(\lambda_{j})-\frac{1}{N}\sum_{k\neq j}^{M}\theta_{2}(\lambda_{j}-\lambda_{k})=2\pi\frac{J_{j}}{N}
\end{equation}
where $\theta_n(\lambda)=2 \arctan(\frac{2\lambda}{n})$ and the Bethe quantum numbers $J_{j}$ are integers for $N+M$ odd and half-odd integers for $N+M$ even. Due to the string structure of the solutions, it is a non-trivial task to determine the set of possible configurations of Bethe quantum numbers. For a pair of complex conjugate roots $\{\lambda,\lambda^\ast\}$ where $\lambda$ is assumed to be located in the upper half part of the complex plane, we proceed to analyse the difference between the corresponding quantum numbers by subtracting the Bethe equations for both conjugate roots. Similar to the spin-$\frac{1}{2}$ case~\cite{2007_Hagemans_JPA_40}, the branch cuts of the inverse tangent of the conjugate root has to be taken into account properly, $\arctan\lambda^\ast=(\arctan\lambda)^\ast \pm \pi$ for $\lambda \in [\mp i,\mp i \infty]$ and $\arctan\lambda^\ast=(\arctan\lambda)^\ast$ elsewhere. Furthermore we restrict to $\text{Re } \lambda \neq 0$ and $\text{Re } (\lambda-\lambda_k) \neq 0 \; \forall k$, implying that the self scattering term between the conjugate roots yields the only non-zero branch cut of the difference between the corresponding Bethe equations, resulting in
\begin{equation}
J_- - J_+ =
\begin{cases}
\;0 \quad \text{if}\;\text{Im } \lambda < \frac{1}{2}, \\
\;1 \quad \text{if}\;\text{Im } \lambda > \frac{1}{2}.
\end{cases}
\label{eq:branchcutdiff}
\end{equation}
More elaborate examples where strings are centered at the origin or additionally have coinciding string centers will be encountered within the low-lying excitation spectrum of the spin-$1$ chain. The behaviour of the quantum numbers in these cases will be treated in section~\ref{sec:deviations}.

Equation~(\ref{eq:branchcutdiff}) exhibits a situation where quantum numbers become equal, invalidating the naively-expected exclusion principle on the quantum numbers for the logarithmic Bethe equations. Usually the approximation of non-deviated strings for the Bethe equations is considered, where the Bethe equations are cast into a more convenient form in terms of the real string centers instead of complex rapidities. The resulting string quantum numbers~$I^n_j$ turn out to be strictly non-repeating, which will allow for a legitimate construction of all the possible combinations of string solutions. However, this approach fails to deal with the effect of string deviations. In order to treat deviations correctly it will be neccessary to reconstruct part of the Bethe quantum numbers~$J_j$ from the string quantum numbers~$I_j$ at a future stage.

The Bethe-Takahashi equations are an adaptation of the Bethe equations for non-deviated string solutions. The basic strategy for their derivation is to take the product over the corresponding Bethe equations for each rapidity inside a string, where the string rapidities in the approximation of non-deviated strings are parametrised as
\begin{equation}
\alpha_{j}^{n,a}=\alpha_{j}^{n}+\frac{i}{2}(n+1-2a)
\label{eq:purestringpar}
\end{equation}
where $\alpha_{j}^{n}\in \mathbb{R}$ is the real part of the string and $a$ specifies the rapidity inside a string of length $n$ labeled by $j$. The product over the free part of the spin-$1$ Bethe equations~(\ref{eq:BEspinS}) for rapidities within one $n$-string gives
\begin{align*}
\prod_{a=1}^{n}\frac{\alpha_{j}^{n,a}-i}{\alpha_{j}^{n,a}+i}&=\prod_{a=1}^{n}\frac{\alpha_{j}^{n}+\frac{1}{2}i(n-1-2a)}{\alpha_{j}^{n}+\frac{1}{2}i(n+3-2a)}=\frac{\alpha_{j}^{n}-\frac{1}{2}i(n+1)}{\alpha_{j}^{n}+\frac{1}{2}i(n+1)} \; \frac{\alpha_{j}^{n}-\frac{1}{2}i(n-1)}{\alpha_{j}^{n}+\frac{1}{2}i(n-1)}\\[1em]
&=\begin{cases}
&\cfrac{\alpha_{j}^{1}-i}{\alpha_{j}^{1}+i}\quad \text{ for } n=1,\\[1.2em]
&\cfrac{1+i\alpha_{j}^{n}\frac{2}{n+1}}{1-i\alpha_{j}^{n}\frac{2}{n+1}}\; \cfrac{1+i\alpha_{j}^{n}\frac{2}{n-1}}{1-i\alpha_{j}^{n}\frac{2}{n-1}}\quad \text{ for } n\ge2.
\end{cases}
\end{align*}

Taking logarithms of this factor to the power $N$ and using the relation between the inverse tangent and logarithm yields for $n=1$
\begin{align*}
\ln \left[\frac{\alpha_{j}^{1}-i}{\alpha_{j}^{1}+i} \right]^{N}&=N \ln \left[-\frac{1+i \alpha_{j}^{1}}{1-i\alpha_{j}^{1}} \right]\\
&=i N \theta_{2}(\alpha_{j}^{1})+i N \pi\mod 2\pi i
\end{align*}
and for $n\ge2$
\begin{align*}
\ln\left[ \frac{1+i\alpha_{j}^{n}\frac{2}{n+1}}{1-i\alpha_{j}^{n}\frac{2}{n+1}}\; \frac{1+i\alpha_{j}^{n}\frac{2}{n-1}}{1-i\alpha_{j}^{n}\frac{2}{n-1}} \right]^{N}
&=i N \theta_{n+1}(\alpha_{j}^{n})+i N \theta_{n-1}(\alpha_{j}^{n}) \mod 2\pi i.
\end{align*}

By rearranging the fractions for strings with $n\ge 2$, a minus sign comes out twice. This implies that there is no $N$ dependence in the parity\footnote{Where parity indicates whether the quantum numbers are integers or half-odd integers.} of the quantum numbers in the end. For one-strings, the minus sign cannot be canceled and is being absorbed into the quantum numbers. This yields a different behaviour of quantum numbers for one-strings compared to other strings of higher length. A difference with respect to the spin-$\frac{1}{2}$ case emerges, where the parity of all string quantum numbers are dependent on both $N$ and $M_n$, irrespective of the string length.

The scattering part of the Bethe equations is not dependent on spin, so this part of the derivation will have the same result as the original spin-$\frac{1}{2}$ Bethe-Takahashi equations~\cite{1972_Takahashi_PTP_48}. The final result for the spin-$1$ Bethe-Takahashi equations is
\begin{equation}
(1-\delta_{n,1})\theta_{n-1}(\alpha_{j}^{n})+\theta_{n+1}(\alpha_{j}^{n})-\frac{1}{N}\sum_{m,k}\Theta_{nm}(\alpha_{j}^{n}-\alpha_{k}^{m})=\frac{2\pi}{N} I_{j}^{n}
\label{eq:mainBTs1}
\end{equation}
where
\begin{equation}
\Theta_{nm}(\alpha)=(1-\delta_{nm})\theta_{|n-m|}(\alpha)+2\theta_{|n-m|+2}(\alpha)+...+2\theta_{n+m-2}(\alpha)+\theta_{n+m}(\alpha)
\end{equation}
and the quantum numbers for one strings $I_{j}^{n=1}$ are integers for $N+M_{1}$ odd and half-odd integers for $N+M_{1}$ even. For higher strings $I_{j}^{n\ge2}$ are integers for $M_{n}$ odd and half-odd integers for $M_{n}$ even.

From the Bethe-Takahashi string quantum numbers the dimensionality of the solutions of a specific configuration of strings can be determined, as the $I_j^n$ are non-repeating. The maximum allowed value of $I_j^n$ is to be determined by placing one of the $n$-strings at infinity. The corresponding quantum number is $I^{n,\infty}$ and is calculated from taking the limit of the Bethe-Takahashi equations for the concerning configuration of strings. The highest possible quantum number for which the corresponding string consists of finite rapidities follows from $I^{n,\infty}$. For the quantum numbers of larger strings, it is necessary to subtract $I^{n,\infty}$ by the length of the string, as each rapidity at infinity that has to be placed back into a string with finite rapidities lowers the maximum allowed quantum number by one. This comment about the limiting string quantum numbers for two or higher string solutions has been made in \cite{1984_Faddeev_JMS_24}. The maximum allowed Bethe-Takahashi quantum number is given by $I^{n,\text{max}}=I^{n,\infty}-n$. 

The number of allowed quantum numbers ranging from $-I^{n,\max}$ until $I^{n,\max}$ is given by $2I^{n,\max}+1$. For a state containing $M_{n}$ $n$-strings, the number of possible distributions of Bethe-Takahashi quantum numbers over $M_{n}$ available $n$-strings is 
\begin{equation}
\prod_n {2I^{n,\max}+1\choose M_{n}}.
\end{equation} 

For the ground state in zero field ($M=N$) consisting of two-strings only, it is straightforward to show that the total number of allowable quantum numbers is $\frac{N}{2}$. Distributing $\frac{N}{2}$ two-strings over this available set of quantum numbers yields only one possible configuration. For the remaining excited states, we must seek a different configuration of string solutions.

Two possible excitations for the spin-$1$ chain were already discussed in~\cite{1982_Takhtajan_PLA_87}, where the ground state sea of two-strings is perturbed with strings of length respectively one and three.  We will extend this reasoning towards higher excitations which yield significant contributions to the dynamical structure factor.

In zero field, we can distinguish between two types of important low lying excitations with different total spin. In the first case, we consider excitations with the same number of rapidities as compared to the ground state, these being in the subsector where $S=0$. Secondly, excitations with important contribution to the dynamical structure factor will have one rapidity removed, thus being the highest weight states in the $S=1$ subsector. For the contributions to the transverse dynamical structure factor $S^{-+}(q,\omega)$ we only need to consider the highest weight states of the $S=1$ sector. The longitudinal structure factor will not directly be interesting in zero field, as $S^{zz}(q,\omega)$ is equal to $S^{-+}(q,\omega)$ up to a factor of two due to the global SU(2) symmetry in this particular case. However, the structure of the important string solutions, including limiting quantum numbers and dimensionality can be investigated for both kinds of excitations and are given in tables~\ref{table:extrans} and~\ref{table:exlong}.

\begin{table}[h!]
\centering
\begin{tabular}{ ccllc }
\toprule
\multirow{2}{*}{2p} & 1 & one-string & $I^{1,\max}=0$ & \multirow{2}{*}{$\dbinom{\frac{N}{2}+1}{\frac{N}{2}-1}$}  \\[0.3em]
 & $\frac{N-2}{2}$ & two-strings &  $I^{2,\max}=\frac{N}{4}$& \\[0.3em]
 \midrule
  \multirow{2}{*}{4p-I} &  $\frac{N-4}{2}$  & two-strings & $I^{2,\max}=\frac{N+2}{4}$ & \multirow{2}{*}{ $3\dbinom{\frac{N}{2}+2}{\frac{N}{2}-2}$ }  \\[0.3em]
 & $1$ & three-string &  $I^{3,\max}=1$& \\[0.3em]
 \midrule
   \multirow{3}{*}{4p-II} &  2  & one-strings & $I^{1,\max}=\frac{1}{2}$ & \multirow{3}{*}{ $3\dbinom{\frac{N}{2}+1}{\frac{N}{2}-3}$ }  \\[0.3em]
  & $\frac{N-6}{2}$ & two-strings &  $I^{2,\max}=\frac{N}{4}$& \\[0.3em]
 & $1$ & three-string &  $I^{3,\max}=1$& \\[0.3em]
 \bottomrule
 \end{tabular}
\caption{Structure of the low-lying excitations for the transverse direction in zero field $M=N-1$. The right column yields the total number of possible solutions of this type of excitation.}
\label{table:extrans}
\end{table}

\begin{table}[h!]
\centering
\begin{tabular}{ ccllc }
\toprule
   \multirow{3}{*}{2p} &  1  & one-string & $I^{1,\max}=0$ & \multirow{3}{*}{ $\dbinom{\frac{N}{2}}{\frac{N}{2}-2}$ }  \\[0.3em]
  & $\frac{N-4}{2}$ & two-strings &  $I^{2,\max}=\frac{N}{4}-\frac{1}{2}$& \\[0.3em]
 & $1$ & three-string &  $I^{3,\max}=0$& \\[0.3em]
  \midrule
 \multirow{2}{*}{4p} & $\frac{N-4}{2}$ & two-strings & $I^{2,\max}=\frac{N}{4}+\frac{1}{2}$ & \multirow{2}{*}{$\dbinom{\frac{N}{2}+2}{\frac{N}{2}-2}$}  \\[0.3em]
 & $1$ & four-string &  $I^{4,\max}=0$& \\[0.3em]
 \bottomrule
 \end{tabular}
\caption{Structure of the low-lying excitations for the longitudinal direction in zero field $M=N$. The right column yields the total number of possible solutions of this type of excitation.}
\label{table:exlong}
\end{table}

For excitations with $S=1$ containing $M=N-1$ rapidities, we can build up the states by breaking up one or more two-strings from the original ground state configuration. One of the rapidities of the destroyed two-string is removed, while the remaining rapidity can only become a real rapidity due to the self conjugacy of the Bethe solutions. The limiting Bethe-Takahashi quantum numbers are computed easily by the aforementioned procedure, leaving space for two holes in the sea of quantum numbers, making this the most elementary two-spinon excitation of the model. The dimensionality of this sector of excitations corresponds to that of two-spinon states in the spin-$\frac{1}{2}$ case. By means of the Thermodynamic Bethe Ansatz, one can retrieve the two-spinon dispersion law of Des Cloizeaux-Pearson type $\epsilon(q)=\frac{\pi}{2} |\sin (q)|$~\cite{1982_Takhtajan_PLA_87}. The results from section~\ref{sec:results} will demonstrate that the matrix elements of the two-spinon excitations will provide for the dominant contribution to the dynamical structure factor.

Higher excitations can be constructed by breaking up an additional two-string. With two removed two-strings and one rapidity placed at infinity, the remaining three rapidities can either be real, or can be used to construct a three-string. For the former, an analysis of the limiting string quantum numbers shows that there are no quantum numbers available, indicating there exist no solutions for three real rapiditites completed with two-strings. For the latter, this string configuration containing a three-string will give rise to four holes in the two-string sea of available quantum numbers. Therefore the string configuration with the presence of a single three-string will be one of the available four-spinon excitations. 

Continuing to the case with three removed two-strings from the original ground state sea, there are a number of string configurations possible with the five remaining rapidities. One of them, with a three-string and two real rapidities completed with a sea of two-strings, gives rise to four holes in the two-string quantum numbers as well, completing the available types of four-spinon configurations. These two variations of four-spinon configurations of strings will lead to the subleading part of the intensity of the transverse dynamical structure factor.

Tables~\ref{table:extrans} and~\ref{table:exlong} might straightforwardly be extended by repeating the line of reasoning of breaking up multiple two-strings from the ground state and placing the rapidities back in various different string configurations. The limiting string quantum numbers provide for the admissibility of the constructed state. The described configurations of strings giving rise to the spinon states in the Babujan-Takhtajan spin-$1$ chain are consistent with the spinon statistics in spin-$s$ chains introduced in~\cite{1998_Frahm_PLA_250} and furthermore form a quasiparticle basis in the thermodynamic limit.

\section{Parametrisation for deviated strings}
\label{sec:deviations}
This section aims to cast the Bethe equations into a numerically solvable set of real equations describing deviated string solutions. The strategy introduced in~\cite{2007_Hagemans_JPA_40} for deviated strings in the spin-$\frac{1}{2}$ Heisenberg model will be applied to the spin-$1$ Bethe equations. In this strategy, the Bethe equations are manipulated and rearranged in such a way that they allow for a convergent iterative numerical solving procedure in order to obtain all string centers and deviations. Our results stated below will be modifications concerning the spin-$1$ chain with respect to the equations obtained in~\cite{2007_Hagemans_JPA_40}.

The most important contributions to the dynamical structure factor will include all two-spinon and four-spinon matrix elements with respect to the ground state. Therefore, the real parametrisation of string deviations shall be derived up to solutions containing a single three-string and an arbitrary number of real rapidities and two-strings.

The two conjugate rapidities building up a deviated two-string can be parametrised as
\begin{equation}
\lambda_{j}^{(2),\pm}=\lambda^{(2)}_{j}\pm \frac{i}{2}\Big(1+2\delta^{(2)}_{j}\Big)
\end{equation}
where $\lambda_j^{(2)}, \delta^{(2)}_j \in \mathbb R$ are respectively the root center and deviation in the imaginary direction from the non-deviated string solution. Furthermore deviated three-string solutions can be parametrised as
\begin{align}
\lambda_{j}^{(3),\pm}&=\lambda^{(3)}_{j}+\epsilon^{(3)}_j\pm i \Big(1+\delta^{(3)}_{j}\Big)\\
\lambda_{j}^{(3),0}&=\lambda^{(3)}_j
\end{align}
where $\lambda_j^{(3)}, \delta^{(3)}_j,\epsilon^{(3)}_j \in \mathbb R$. Here, a deformation~$\epsilon_j^{(3)}$ of the real part of the outermost rapidities with respect to the string center~$\lambda_j^{(3)}$ must be considered as well, as rapidities are only constrained to be complex conjugate pairs.

Various expressions for the parameters for deformed string solutions can be extracted by adding and subtracting logarithmic Bethe equations of two conjugate roots. A careful treatment of the branch cuts of the addition of two inverse tangents with complex conjugate arguments is essential,
\begin{equation}
\arctan(a+i b)+\arctan(a-i b)=\xi(a,1+b)+\xi(a,1-b)
\label{eq:sumatan}
\end{equation}
where $a,b \in \mathbb{R}$ and $\xi(\epsilon,\delta)$ is defined as
\begin{equation}
\xi(\epsilon,\delta)=\arctan\left(\frac{\epsilon}{\delta} \right)+\pi\;\Theta(-\delta)\;\text{sign}(\epsilon).
\end{equation}
The conventions of values of the step functions evaluated at zero are $\text{sign}(0)=0$ and $\Theta(0)=\frac{1}{2}$, where $\Theta(\delta)$ denotes the Heaviside step function. Another important limit is given by $\lim_{\delta \rightarrow 0} \xi(\epsilon,\delta)=\frac{\pi}{2} \text{sign} (\epsilon)$.

The final parametrised expression for the Bethe equations of one-string rapidities $\lambda_{j}^{(1)}$ is obtained by applying equation~(\ref{eq:sumatan}) to the scattering terms of a one-string with two conjugate roots inside a higher string,
\begin{align}
\arctan\Big(\lambda_{j}^{(1)}\Big)&=\frac{\pi}{N}J_{j}^{(1)}+\frac{1}{N}\sum_{k\neq j}^{(n=1)}\arctan\Big(\lambda_{j}^{(1)}-\lambda_{k}^{(1)}\Big)\nonumber\\
+&\frac{1}{N}\sum_{k}^{(n=2)}\Big[\xi\Big(\lambda_{j}^{(1)}-\lambda^{(2)}_{k},\frac{3}{2}+\delta^{(2)}_{k}\Big)+\xi\Big(\lambda_{j}^{(1)}-\lambda^{(2)}_{k},\frac{1}{2}-\delta^{(2)}_{k}\Big)\Big]\nonumber\\
+&\frac{1}{N}\sum_{k}^{(n=3)}\Big[\xi\Big(\lambda_{j}^{(1)}-\lambda^{(3)}_{k}-\epsilon^{(3)}_k,2+\delta^{(3)}_{k}\Big)+\xi\Big(\lambda_{j}^{(1)}-\lambda^{(3)}_{k}-\epsilon^{(3)}_k,-\delta^{(3)}_{k}\Big)\nonumber\\
&+\arctan\Big(\lambda_j^{(1)}-\lambda_k^{(3)} \Big)\Big].
\label{eq:parameterisation_one_strings}
\end{align}
An important step is to link the set of string quantum numbers of the Bethe-Takahashi equations $I^{(n)}_j$ to the quantum numbers of the original Bethe equations $J^{(n)}_j$ used in the parametrisation for deviated strings. This can in general be done by taking the limit of vanishing deviations $\delta,\epsilon \rightarrow 0$ and inserting the Bethe-Takahashi equations subsequently. In this limit, equation~(\ref{eq:parameterisation_one_strings}) becomes
\begin{equation}
J_j^{(1)}=I_j^{(1)}-\frac{1}{2}\sum_k^{(n=3)}\text{sign}\Big(\lambda_j^{(1)}-\lambda_k^{(3)}\Big).
\end{equation}

In order to obtain an equation for the two-string root center $\lambda^{(2)}_{j}$, the corresponding logarithmic Bethe equations for the two roots inside a two-string must be added properly according to equation~(\ref{eq:sumatan}). This identity is to be applied to both the sum of the free parts of the two Bethe equations as well as all the sums over conjugate roots within the scattering parts. The result is
\begin{align}
\nonumber \xi\Big(\lambda^{(2)}_{j},&\frac{3}{2}+\delta^{(2)}_{j}\Big)+\xi\Big(\lambda^{(2)}_{j},\frac{1}{2}-\delta^{(2)}_{j}\Big)=\frac{\pi}{N}\Big(J_{+}^{(2)}+J_{-}^{(2)}\Big)\\ \nonumber
+&\frac{1}{N}\sum_{k}^{(n=1)}\Big[\xi\Big(\lambda^{(2)}_{j}-\lambda_{k}^{(1)},\frac{3}{2}+\delta^{(2)}_{j}\Big)+\xi\Big(\lambda^{(2)}_{j}-\lambda_{k}^{(1)},\frac{1}{2}-\delta^{(2)}_{j}\Big)\Big]\\ \nonumber
+&\frac{1}{N}\sum_{k\neq j}^{(n=2)}\Big[\xi\Big(\lambda^{(2)}_{j}-\lambda^{(2)}_{k},\delta^{(2)}_{j}+\delta^{(2)}_{k}+2\Big)+\xi\Big(\lambda^{(2)}_{j}-\lambda^{(2)}_{k},-\delta^{(2)}_{j}-\delta^{(2)}_{k}\Big)\\
&+\xi\Big(\lambda^{(2)}_{j}-\lambda^{(2)}_{k},\delta^{(2)}_{j}-\delta^{(2)}_{k}+1\Big)+\xi\Big(\lambda^{(2)}_{j}-\lambda^{(2)}_{k},-\delta^{(2)}_{j}+\delta^{(2)}_{k}+1\Big)\Big]\nonumber\\
+&\frac{1}{N}\sum_{k}^{(n=3)}\Big[\xi\Big(\lambda^{(2)}_{j}-\lambda^{(3)}_{k},\frac{3}{2}+\delta^{(2)}_{j}\Big)+\xi\Big(\lambda^{(2)}_{j}-\lambda^{(3)}_{k},\frac{1}{2}-\delta^{(2)}_{j}\Big)\nonumber\\
&+\xi\Big(\lambda^{(2)}_{j}-\lambda^{(3)}_{k}-\epsilon^{(3)}_k,\frac{1}{2}+\delta^{(2)}_{j}-\delta^{(3)}_{k}\Big)+\xi\Big(\lambda^{(2)}_{j}-\lambda^{(3)}_{k}-\epsilon^{(3)}_k,\frac{3}{2}-\delta^{(2)}_{j}+\delta^{(3)}_{k}\Big)\nonumber\\
&+\xi\Big(\lambda^{(2)}_{j}-\lambda^{(3)}_{k}-\epsilon^{(3)}_k,\frac{5}{2}+\delta^{(2)}_{j}+\delta^{(3)}_{k}\Big)+\xi\Big(\lambda^{(2)}_{j}-\lambda^{(3)}_{k}-\epsilon^{(3)}_k,-\frac{1}{2}-\delta^{(2)}_{j}-\delta^{(3)}_{k}\Big)\Big].
\end{align}

The determination of Bethe quantum numbers from string quantum numbers is again to be derived from the limit with zero deviations and plugging in Bethe-Takahashi equations,
\begin{equation}
J_{+}^{(2)}+J_{-}^{(2)}=I_{j}^{(2)}-\frac{1}{2}\sum_{k\neq j}^{(n=2)}\text{sign}\left(\lambda^{(2)}_{j}-\lambda^{(2)}_{k}\right)-\sum_{k}^{(n=3)}\text{sign}\left(\lambda^{(2)}_{j}-\lambda^{(3)}_{k}\right).
\label{eq:qnolinktwo}
\end{equation}

The equation for two-string deviation $\delta^{(2)}_{j}$ might be considered by taking the difference between the Bethe equations of conjugate roots. It is however both equivalent and more convenient to take the quotient of the original Bethe equations in product form,
\begin{align}
\left[\frac{1+\delta^{(2)}_{j}}{\delta^{(2)}_{j}}\right]^{2}&=\left[\frac{{(\lambda^{(2)}_{j})}^{2}+(\delta^{(2)}_{j}+2)(\delta^{(2)}_{j}+1)+\frac{1}{4}}{{(\lambda^{(2)}_{j})}^{2}+\delta^{(2)}_{j}(\delta^{(2)}_{j}-1)+\frac{1}{4}} \right]^{N}\; \prod_{k}^{(n=1)}\frac{{(\delta^{(2)}_{j}-\frac{1}{2})}^{2}+{(\lambda^{(2)}_{j}-\lambda_{k}^{(1)})}^{2}}{{(\delta^{(2)}_{j}+\frac{3}{2})}^{2}+{(\lambda^{(2)}_{j}-\lambda_{k}^{(1)})}^{2}}\nonumber \\
& \cdot \prod_{k\neq j}^{(n=2)}\frac{{(\delta^{(2)}_{j}+\delta^{(2)}_{k})}^{2}+{(\lambda^{(2)}_{j}-\lambda^{(2)}_{k})}^{2}}{{{(\delta^{(2)}_{j}+\delta^{(2)}_{k}+2)}}^{2}+{{(\lambda^{(2)}_{j}-\lambda^{(2)}_{k})}}^{2}} \; \frac{{(1-\delta^{(2)}_{j}+\delta^{(2)}_{k})}^{2}+{(\lambda^{(2)}_{j}-\lambda^{(2)}_{k})}^{2}}{{(1+\delta^{(2)}_{j}-\delta^{(2)}_{k})}^{2}+{(\lambda^{(2)}_{j}-\lambda^{(2)}_{k})}^{2}}\nonumber\\
& \cdot \prod_{k}^{(n=3)}\frac{{(\frac{1}{2}-\delta^{(2)}_{j})}^{2}+{(\lambda^{(2)}_{j}-\lambda^{(3)}_{k})}^{2}}{{(\frac{3}{2}+\delta^{(2)}_{j})}^{2}+{(\lambda^{(2)}_{j}-\lambda^{(3)}_{k})}^{2}} \; \frac{{(\frac{1}{2}+\delta^{(2)}_{j}+\delta^{(3)}_k)}^{2}+{(\lambda^{(2)}_{j}-\lambda^{(3)}_{k}-\epsilon^{(3)}_k)}^{2}}{{(\frac{5}{2}+\delta^{(2)}_{j}+\delta^{(3)}_k)}^{2}+{(\lambda^{(2)}_{j}-\lambda^{(3)}_{k}-\epsilon^{(3)}_k)}^{2}}\nonumber \\
&\phantom{\prod_{k}^{(n=3)}} \cdot \frac{{(\frac{3}{2}-\delta^{(2)}_{j}+\delta^{(3)}_k)}^{2}+{(\lambda^{(2)}_{j}-\lambda^{(3)}_{k}-\epsilon^{(3)}_k)}^{2}}{{(\frac{1}{2}+\delta^{(2)}_{j}-\delta^{(3)}_k)}^{2}+{(\lambda^{(2)}_{j}-\lambda^{(3)}_{k}-\epsilon^{(3)}_k)}^{2}}.
\label{eq:devtwostring}
\end{align}

The square on the left-hand side of equation~(\ref{eq:devtwostring}) leaves the sign of the two-string deviation undetermined. However, from equation~(\ref{eq:branchcutdiff}) where the two-string is not centered at zero,
\begin{align}
J_{-}^{(2)}-J_{+}^{(2)}&=\Theta\big(\delta^{(2)}_j\big). \label{eq:twostringnarrowwideheaviside}
\end{align}
Together with the link between the string quantum numbers and Bethe quantum numbers for deviated two-strings in equation~(\ref{eq:qnolinktwo}) and the parity of the quantum numbers, the sign of $\delta^{(2)}_j$ can be fixed via the Heaviside function in equation~(\ref{eq:twostringnarrowwideheaviside}). The correct sequence of the two-strings with respect to the three-strings is however an unavoidable part of  equation~(\ref{eq:qnolinktwo}) by the presence of the step functions. This can be provided by iteratively solving the Bethe-Takahashi equations for the non-deviated string centers and using the results as initial values for the determination of the sign of the two-string deviations $\delta^{(2)}_j$.

For symmetric distributions of string quantum numbers and depending on the parity of these quantum numbers, multiple strings could be centered at the origin. Therefore we must extend the reasoning of the determination of the sign of deviations and Bethe quantum numbers described in the previous paragraph to these cases.  For the situations we take under consideration, we might deal with strings centered at zero for the ground state or two-spinon states.

By the subtraction of Bethe equations of conjugate roots performed to obtain equation~(\ref{eq:branchcutdiff}), we now consider the case where $\text{Re } \lambda=0$, implying $\lambda= - \lambda^\ast$. Still assuming non-coinciding root centers, the self-scattering term between the conjugate roots again yields a branch cut at $\text{Im } \lambda = \frac{1}{2}$, while the subtraction between the free terms yields a branch cut at $\text{Im } \lambda = 1$ for the spin-$1$ chain in particular. It might then be concluded that equation~(\ref{eq:twostringnarrowwideheaviside}) safely holds for two-strings at zero, assuming that $|\delta^{(2)}_j| < \frac{1}{2}$.  Furthermore, symmetric two-spinon states have a two-string as well as a real-rapidity centered at zero. In this situation, only 
the difference between one-string scattering terms needs to be additionally taken into consideration. This case only yields extra step behaviour at $\text{Im } \lambda = 1$, which once more will not provide implications on the algorithm.

The sum of the Bethe equations for all roots within a three-string yields an expression for the root center of a three-string $\lambda_j^{(3)}$,
\begin{align}
\arctan\Big(&\lambda_j^{(3)}\Big)+\xi\Big(\lambda_j^{(3)}+\epsilon_j^{(3)},2+\delta_j^{(3)}\Big)+\xi\Big(\lambda_j^{(3)}+\epsilon_j^{(3)},-\delta_j^{(3)}\Big)=\frac{\pi}{N}\Big(J_-^{(3)}+J_0^{(3)}+J_+^{(3)}\Big)\nonumber\\
+&\frac{1}{N}\sum_k^{(n=1)} \Big[\arctan\Big(\lambda_j^{(3)}-\lambda_k^{(1)}\Big)+\xi\Big(\lambda_j^{(3)}+\epsilon_j^{(3)}-\lambda_k^{(1)},2+\delta_j^{(3)}\Big)\nonumber\\
&\phantom{}+\xi\Big(\lambda_j^{(3)}+\epsilon_j^{(3)}-\lambda_k^{(1)},-\delta_j^{(3)}\Big)\Big]\nonumber\\
+&\frac{1}{N}\sum_k^{(n=2)}\Big[
\xi\Big(\lambda_j^{(3)}-\lambda_k^{(2)},\frac{3}{2}+\delta_k^{(2)}\Big)+\xi\Big(\lambda_j^{(3)}-\lambda_k^{(2)},\frac{1}{2}-\delta_k^{(2)}\Big)\nonumber\\
&\phantom{}+\xi\Big(\lambda_j^{(3)}+\epsilon_j^{(3)}-\lambda_k^{(2)},\frac{5}{2}+\delta_j^{(3)}+\delta_k^{(2)}\Big)+\xi\Big(\lambda_j^{(3)}+\epsilon_j^{(3)}-\lambda_k^{(2)},-\frac{1}{2}-\delta_j^{(3)}-\delta_k^{(2)}\Big)\nonumber\\
&\phantom{}+\xi\Big(\lambda_j^{(3)}+\epsilon_j^{(3)}-\lambda_k^{(2)},\frac{3}{2}+\delta_j^{(3)}-\delta_k^{(2)}\Big)+\xi\Big(\lambda_j^{(3)}+\epsilon_j^{(3)}-\lambda_k^{(2)},\frac{1}{2}-\delta_j^{(3)}+\delta_k^{(2)}\Big) \Big].
\end{align}
Again taking the limit of vanishing deviations and substituting in the Bethe-Takahashi equations, we obtain a relation between the three-string quantum numbers,
\begin{align}
J^{(3)}_++J^{(3)}_0+J^{(3)}_-&=I^{(3)}_j-\frac{1}{2}\sum_k^{(n=1)}\text{sign}\Big(\lambda_j^{(3)}-\lambda_k^{(1)}\Big)-\sum_k^{(n=2)}\text{sign}\Big(\lambda_j^{(3)}-\lambda_k^{(2)}\Big)+\frac{N}{2}\text{sign}\Big(\lambda_j^{(3)}\Big).
\end{align}

The remaining three-string deviations $\delta^{(3)}_j$ and $\epsilon^{(3)}_j$ can be found in the following way. We will consider the Bethe equations of the outermost complex conjugate roots of the three-string. The quotient of the corresponding Bethe equations results in the modulus squared of $\delta^{(3)}_j$ and $\epsilon^{(3)}_j$, while the sum of the logarithmic Bethe equations yields the argument. The quotient of the Bethe equations for $\lambda^{(3),+}_j$ and $\lambda^{(3),-}_j$ is
\begin{align}
{(\delta_j^{(3)})}^2&+{(\epsilon_j^{(3)})}^2={(r_j)}^2=\Big[{(2+\delta_j^{(3)})}^2+{(\epsilon_j^{(3)})}^2\Big]\left[ \frac{3+2\delta_j^{(3)}}{1+2\delta_j^{(3)}} \right]^2\left[\frac{{(\lambda_j^{(3)}+\epsilon_j^{(3)})}^2+{(\delta_j^{(3)})}^2}{{(\lambda_j^{(3)}+\epsilon_j^{(3)})}^2+{(2+\delta_j^{(3)})}^2}  \right]^N\nonumber\\
&\cdot \prod_k^{(n=1)} \frac{{(\lambda_j^{(3)}+\epsilon_j^{(3)}-\lambda_k^{(1)})}^2+{(2+\delta_j^{(3)})}^2}{{(\lambda_j^{(3)}+\epsilon_j^{(3)}-\lambda_k^{(1)})}^2+{(\delta_j^{(3)})}^2}\nonumber\\
&\cdot \prod_k^{(n=2)} \frac{{(\lambda_j^{(3)}+\epsilon_j^{(3)}-\lambda_k^{(2)})}^2+{(\frac{5}{2}+\delta_j^{(3)}+\delta_k^{(2)})}^2}{{(\lambda_j^{(3)}+\epsilon_j^{(3)}-\lambda_k^{(2)})}^2+{(\frac{1}{2}+\delta_j^{(3)}+\delta_k^{(2)})}^2}\;\frac{{(\lambda_j^{(3)}+\epsilon_j^{(3)}-\lambda_k^{(2)})}^2+{(\frac{3}{2}+\delta_j^{(3)}-\delta_k^{(2)})}^2}{{(\lambda_j^{(3)}+\epsilon_j^{(3)}-\lambda_k^{(2)})}^2+{(\frac{1}{2}-\delta_j^{(3)}+\delta_k^{(2)})}^2},
\label{eq:threestringmodulus}
\end{align}
while the sum of the logarithmic Bethe equations of $\lambda^{(3),+}_j$ and $\lambda^{(3),-}_j$ results in 
\begin{align}
\xi&\Big(\epsilon_j^{(3)},-\delta_j^{(3)}\Big)=\theta_j=-\xi\Big(\epsilon_j^{(3)},2+\delta_j^{(3)}\Big)-\pi\Big(J^{(3)}_++J^{(3)}_-\Big)\nonumber\\
&+N\left[\xi\Big(\lambda_j^{(3)}+\epsilon_j^{(3)},2+\delta_j^{(3)}\Big)+\xi\Big(\lambda_j^{(3)}+\epsilon_j^{(3)},-\delta_j^{(3)})\right]\nonumber\\
&-\sum_k^{(n=1)} \Big[\xi\Big(\lambda_j^{(3)}+\epsilon_j^{(3)}-\lambda_k^{(1)},2+\delta_j^{(3)}\Big)+\xi\Big(\lambda_j^{(3)}+\epsilon_j^{(3)}-\lambda_k^{(1)},-\delta_j^{(3)}\Big)\Big]\nonumber\\
&-\sum_k^{(n=2)}\Big[\xi\Big(\lambda_j^{(3)}+\epsilon_j^{(3)}-\lambda_k^{(2)},\frac{5}{2}+\delta_j^{(3)}+\delta_k^{(2)}\Big)+\xi\Big(\lambda_j^{(3)}+\epsilon_j^{(3)}-\lambda_k^{(2)},-\frac{1}{2}-\delta_j^{(3)}-\delta_k^{(2)}\Big)\nonumber\\
&\phantom{xxx}+\xi\Big(\lambda_j^{(3)}+\epsilon_j^{(3)}-\lambda_k^{(2)},\frac{3}{2}+\delta_j^{(3)}-\delta_k^{(2)}\Big)+\xi\Big(\lambda_j^{(3)}+\epsilon_j^{(3)}-\lambda_k^{(2)},\frac{1}{2}-\delta_j^{(3)}+\delta_k^{(2)}\Big) \Big].
\label{eq:sumlogthreearg}
\end{align}
The values for the three-string deviations are instantly extracted from the modulus and argument
\begin{align}
\delta_j^{(3)}&=-|r_j| \cos\theta_j\\
\epsilon_j^{(3)}&=|r_j| \sin\theta_j
\end{align}

At this moment, the sum $J^{(3)}_++J^{(3)}_-$ appearing in equation~(\ref{eq:sumlogthreearg}) remains undetermined. However, while the parity of the quantum numbers is already known, equation~(\ref{eq:branchcutdiff}) yields $J^{(3)}_--J^{(3)}_+=1$ provided that $\delta^{(3)}_j > -\frac{1}{2}$ and $\lambda^{(3)}_j \neq 0$.  The former and the latter property fixes the evenness or oddness of $J^{(3)}_++J^{(3)}_-$, while the argument is defined modulo $2\pi$. It is therefore sufficient to only determine whether the sum of the two quantum numbers is even or odd. 

Special attention needs to be given to the behaviour of the deviations of three-strings for symmetric distributions of all string quantum numbers. We focus merely on the states containing one three-string at most as they are of interest for the four-spinon states. For a three-string centered at zero and a symmetric distribution of the remaining quantum numbers of other strings, the deviation along the real axis $\epsilon^{(3)}_j$ must vanish due to the symmetry, while equation~(\ref{eq:threestringmodulus}) allows $\delta^{(3)}_j=0$ simultaneously. In this case the Bethe equations for the three-string become singular. The existence of such singular solutions is similar to the spin-$\frac{1}{2}$ case, where the Bethe equations however become singular at $\lambda=\pm \frac{1}{2}i$. It has been argued analytically~\cite{2007_Hagemans_JPA_40} that the matrix elements of singular states must vanish. Our spin-$1$ results will provide numerical evidence for this statement. 

Away from the singular solution at zero root center, but still for small deviations, ${( r_j )}^2$ in equation~(\ref{eq:threestringmodulus}) becomes exponentially small in system size. This scaling implies that for small root centers, three-string deviations are exponentially suppressed with system size. For root centers far away from zero, the three-string deviations remain large. The size of the deviations will yield numerical problems in the evaluation of the matrix elements exponentially close to the singularities. In general this difficulty is overcome in the spin-$\frac{1}{2}$ case by regularising the matrix element expressions. With large two-string deviations and either small or large three-string deviations within a Bethe state, the regularisation of the different behaviour for the strings in this case needs to be treated with special attention in section~\ref{sec:correlationfunctions}.

Another special class of solutions emerges among the two-spinon excitations in the $S=0$ subsector. A symmetric distribution of quantum numbers in this case with a three-string and a single one-string located at zero, provides a problem for the singular three-string solution at zero. Coinciding rapitidies in general lead to vanishing Bethe vectors, but whenever different string quantum numbers coincide at zero, string deviations usually  regularise these cases. However, in this case the three-string deviations vanish, and the one-string and real rapidity from the three-string coincide, forming an exceptional solution to the Bethe equations. It has been argued that this class of solutions, however, yields a non-zero Bethe vector~\cite{1986_Avdeev_TMP_69}. We will again provide numerical evidence that the matrix elements of these exceptional singular states vanish.

At this stage, all required initial conditions for a convergent algorithm on the Bethe equations including string deviations are set. The Bethe equations are parametrised for string centers and deviations in such a way that they are amenable to an iterative algorithm. The numerical strategy might be summarised as follows. We start off with a state defined from the Bethe-Takahashi string quantum numbers, which form the starting position to solve the Bethe-Takahashi equations iteratively. These non-deviated string solutions set the initial conditions and the signs of the two-string deviations for the succeeding iterative procedure, where the full parametrisation of the Bethe equations in terms of string centers and deviations is to be solved.
\begin{figure}[t!]
\centering
\includegraphics{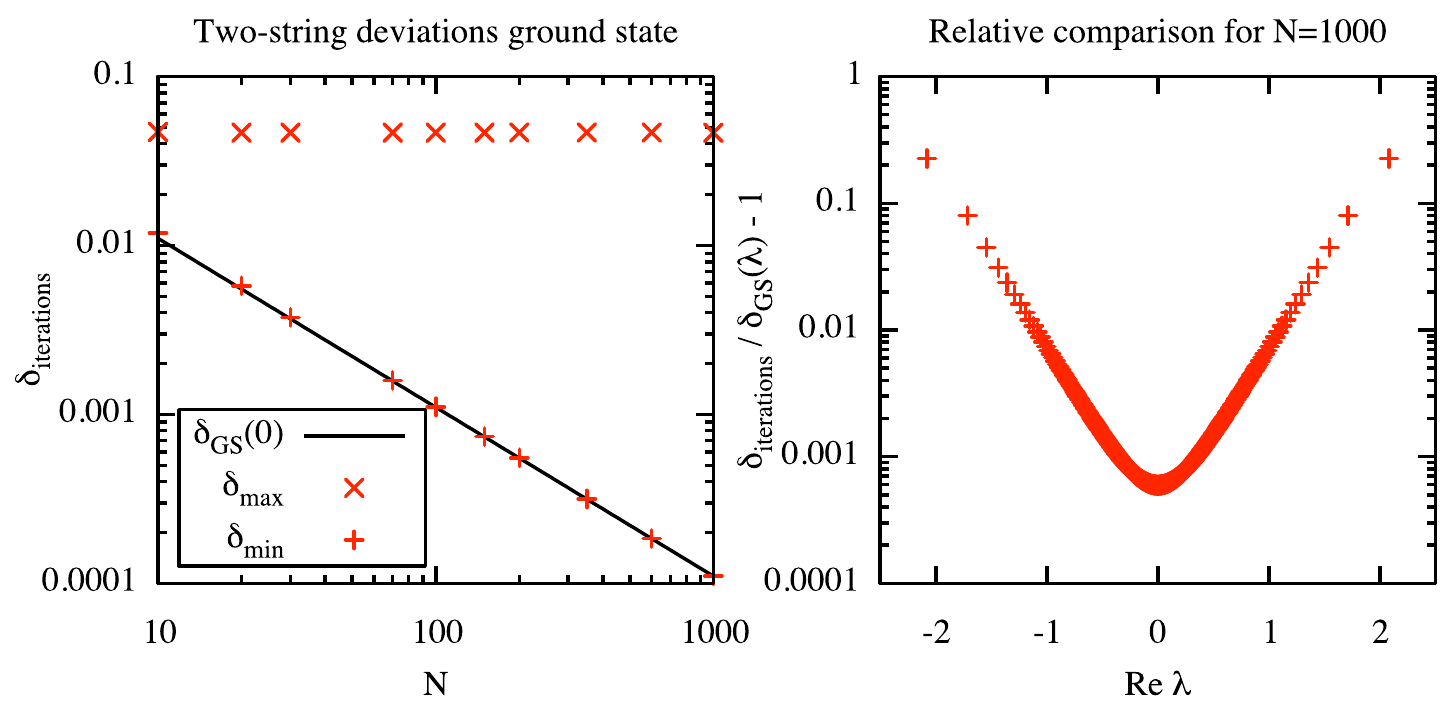}
\caption{Left: Two-string deviations for the ground state for the innermost ($\delta_\text{min}$) and outermost ($\delta_\text{max}$) two-string respectively, obtained by the iterative procedure on string deviations. The solid line corresponds to $\delta_\text{GS}(\lambda)|_{\lambda=0}$ from equation~(\ref{eq:devega}). Right:~Relative comparison of two-string deviations between results from numerics and analytic predictions.}
\label{fig:gsdevcompare}
\end{figure}

The results of the iterative procedure might be compared to previous analytic predictions for the two-string deviations in the ground state. The first available analytical method is based on an adaptation of the Euler-Maclaurin formula applied to the Thermodynamic Bethe Ansatz for spin-$1$~\cite{1990_deVega_JPA_23}, while the second method is a result of nonlinear integral equations for the spin-$1$ ground state~\cite{1990_Kluemper_JPA_23}. Both approaches give an identical prediction up to order $1/N$ for the two-string deviations of the ground state as a function of the string center $\lambda$,
\begin{equation}
\delta_\text{GS}(\lambda) = \frac{\ln2}{4\pi} \frac{1}{N \sigma(\lambda)} =\frac{\ln2}{2\pi N} \cosh(\pi \lambda)
\label{eq:devega}
\end{equation}
where $\sigma(\lambda)$ is the density of two-strings in the thermodynamic limit.

Figure~\ref{fig:gsdevcompare} (left) shows the results on deviations of the innermost and outermost two-string as a function of system size. For the innermost two-string, the data reflects the $1/N$ behaviour of the deviations. However, the outermost two-strings show persistingly constant deviations at $\delta=0.0466$ independent of system size. This constant deviation corresponds to the value found in~\cite{1987_Avdeev_TMP_71}.

Figure~\ref{fig:gsdevcompare} (right) makes a relative comparison between the ground state deviations obtained by our numerical method and equation~(\ref{eq:devega}) as a function of the root center $\lambda$ for fixed system size. Only the bulk of the two-strings shows a good equivalence to equation~(\ref{eq:devega}), however there still remains a significant difference. 

For short chains, rapidities of all states including string deviations are stated in appendix~\ref{app:tables} as a result of the numerical method described in this section. For the energies of all the Bethe states, we find perfect agreement with exact diagonalisation of the Hamiltonian for small system sizes. For larger system sizes, the algorithm described throughout this section enables us to solve the Bethe equations for the complete spectrum of two-spinon and four-spinon states. These results will be used in the computation of the dynamical structure factor in section~\ref{sec:results}.

\section{Matrix elements}
\label{sec:correlationfunctions}

Our method relies on the matrix element expressions for higher spin chains obtained in~\cite{2007_Castro_Alvaredo_JPA_40},  which is based on a combination of the fusion of R-matrices, the inverse scattering method and Slavnov's theorem \cite{1989_Slavnov_TMP_79,1990_Slavnov_TMP_82}. We will adapt the result for general spin-$s$ chains to a formula applicable to our numerical procedure and set some conventions and notations along the way, to obtain an expression for the matrix elements $|F^a_{ q}|^2=| \langle \text{GS} | S^a_q | \alpha \rangle |^2$ in equation~(\ref{eq:defDSF2})  for $a=\pm,z$.

From~\cite{2007_Castro_Alvaredo_JPA_40} we have for the transverse matrix elements
\begin{align}
F_j^+ (\{\lambda\},\{\mu\})&=\langle \psi (\{\lambda\}) | S_j^+ | \psi(\{\mu\})\rangle\\
&= \frac{\varphi_{j-1}(\{\lambda\})\varphi_j (\{\lambda\})}{\varphi_{j}(\{\mu\})\varphi_{j-1} (\{\mu\})} F^-_j (\{\mu\},\{\lambda\})\\
&= \frac{\varphi_{j-1}(\{\lambda\})}{\varphi_{j-1}(\{\mu\})} \frac{ \prod_k^{l+1} (\mu_k-s_j \eta)}{ \prod_k^{l} (\lambda_k-s_j \eta)} \frac{ \det \mathcal C}{\prod_{i<j} (\mu_j-\mu_i) \prod_{i<j} (\lambda_i-\lambda_j)}
\end{align}
where
\begin{align}
\mathcal C_{ab} &= H_{ab}(\{\mu\},\{\lambda\}) \quad \text{for } b\neq l+1\\
\mathcal C_{a,l+1} &=-i \partial_{\mu_a} p^{s_j}(\mu_a)\\
p^{(s_j)} (\lambda) &= i \log \left[ \frac{\lambda - s_j \eta}{\lambda+s_j \eta}\right] \\
H_{ab}&=\frac{\eta}{\mu_a-\lambda_b}\left[ \prod_{j\neq a} (\mu_j-\lambda_b+\eta)- d(\lambda_b) \prod_{j\neq a}(\mu_j-\lambda_b-\eta) \right]\\
d(\lambda)&=\prod_j \frac{\lambda-s_j \eta}{\lambda+s_j \eta}.
\end{align}

Furthermore, as the Bethe states are not normalised, we need to divide by the norms of  $| \psi (\{\lambda\}) \rangle$ and $| \psi (\{\mu\}) \rangle$, which are given by the Gaudin determinant,
\begin{align}
\langle \psi (\{\lambda\}) | \psi (\{\lambda\}) \rangle& = \eta^l \prod_{a\neq b} \frac{\lambda_a-\lambda_b+\eta}{\lambda_a-\lambda_b} \det \Phi(\{\lambda\})\\
\Phi_{ab}(\{\lambda\})=\;&\delta_{ab}\left[N\partial_{\lambda_a}\theta_2(\lambda_a)-\sum_{k\neq a}\partial_{\lambda_a}\theta_2(\lambda_a-\lambda_k)\right]\nonumber \\
&+(1-\delta_{ab})\partial_{\lambda_a}\theta_2(\lambda_a-\lambda_b).
\end{align}
We will adopt conventions $\eta=i$, $M=l+1$, $\varphi_j(\{\lambda\})=e^{-iP_\lambda j}$ and define $\phi_n(\lambda)=\lambda+\frac{i n}{2}$. In order to obtain the final result we take the Fourier transform $S^a_q=\frac{1}{\sqrt{N}}\sum_{j=1}^N e^{iqj} S_j^a$. The matrix elements of the Fourier transformed operators are
\begin{align}
|F^-_q|^2=N &\delta_{q,q_\lambda-q_\mu} \prod_{j=1}^M |\phi_{-2}(\mu_j)|^2 \prod_{j=1}^{M-1} |\phi_{-2}(\lambda_j)|^{-2}
\prod_{j,k}^{j\neq k} |\phi_{2}(\mu_j-\mu_k)|^{-1} \nonumber\\
&\cdot \prod_{j,k}^{j\neq k} |\phi_{2}(\lambda_j-\lambda_k)|^{-1} \frac{|\det H^- (\{\mu\},\{\lambda\})|^2}{||\{\mu\}|| \;||\{\lambda\}||}.
\end{align}
where
\begin{equation}
||\{\lambda\}||=|\det \Phi_{ab}(\{\lambda\})|
\end{equation}
\begin{equation}
H^-_{ab}(\{\mu\},\{\lambda\})=
\begin{cases}
	 & H_{ab}(\{\mu\},\{\lambda\}) \quad \text{for } b<M\\
	 & \cfrac{2}{\phi_2(\mu_a) \phi_{-2}(\mu_a)} \quad \text{for } b=M
\end{cases}
\end{equation}
\begin{equation}
H_{ab}(\{\mu\},\{\lambda\})=\frac{1}{\phi_0(\mu_a-\lambda_b)}\left[ \prod_{j\neq a}\phi_2(\mu_j-\lambda_b)-\left[\frac{\phi_{-2}(\lambda_b)}{\phi_{2}(\lambda_b)}\right]^N\prod_{j\neq a}\phi_{-2}(\mu_j-\lambda_b)\right].
\end{equation}

The result for longitudinal matrix elements for chains of spin-$s$ is~\cite{2007_Castro_Alvaredo_JPA_40}
\begin{equation}
F^z_j (\{\lambda\},\{\mu\})=\frac{\varphi_j(\{\lambda\})}{\varphi_j(\{\mu\})} \frac{s_j \det H-\sum_{p=1}^l\prod_{k=1}^l (\lambda_k-\lambda_p+\eta)\det \mathcal Z}{\prod_{j<j} (\mu_i-\mu_j)(\lambda_j-\lambda_i)}
\end{equation}
where
\begin{align}
\mathcal Z_{ab}^{(p)} &= H_{ab}(\{\mu\},\{\lambda\}) \quad \text{for } b \neq p\\
\mathcal Z_{ap}^{(p)} &=-i \partial_a p^{(s_j)}(\mu_a) \prod_{k=1}^l \frac{\mu_k+s_j \eta}{\lambda_k+s_j \eta}.
\end{align}
Taking the Fourier transform and dividing by the norms of the Bethe states, the expression for the matrix elements ready to be used in computations for the spin-$1$ case becomes
\begin{align}
|F^z_q|^2=N &\delta_{q,q_\lambda-q_\mu} \prod_{j,k}^{j\neq k} |\phi_{2}(\mu_j-\mu_k)|^{-1} \prod_{j,k}^{j\neq k} |\phi_{2}(\lambda_j-\lambda_k)|^{-1} \nonumber\\
&\cdot \frac{|\det H (\{\lambda\},\{\mu\})-\sum_{p}^M\prod_{k}^M \phi_2 (\lambda_k-\lambda_p)\det \mathcal Z^{(p)}|^2}{||\{\mu\}|| \;||\{\lambda\}||}
\end{align}
where
\begin{align}
\mathcal Z_{ab}^{(p)} &= H_{ab}(\{\mu\},\{\lambda\}) \quad \text{for } b \neq p\\
\mathcal Z_{ap}^{(p)} &= \frac{2}{\phi_2(\mu_a)\phi_{-2}(\mu_a)} \prod_{k=1}^l \frac{\phi_2(\mu_k)}{\phi_2(\lambda_k)}.
\end{align}

For the spin-1 chain as elaborated in section~\ref{sec:deviations}, we encounter Bethe states with exponentially small deviations for the three-string, while the remaining two-strings still have fairly large deviations. A straightforward generalisation of the reduced determinants for the spin-$\frac{1}{2}$ chain~\cite{2005_Caux_JSTAT_P09003} is not sufficient, as the effect of the remaining important deviations would be neglected. We aim to remove singularities from the determinants and prefactors present in the matrix element expressions, while keeping track of the effect of the algebraically large deviations.

The singularities present in the norm of the Bethe states can be extracted from the Gaudin determinant as follows. The Gaudin matrix can be written as
\begin{equation}
\Phi_{ab}=\delta_{ab}\left[ d_a - \sum_{k\neq a} o_{ak} \right]+(1-\delta_{ab}) o_{ab}
\end{equation}
where
\begin{align}
d_a&=N\partial_{\lambda_a}\theta_2(\lambda_a) \\
o_{ab}&=\partial_{\lambda_a}\theta_2(\lambda_a-\lambda_b).
\end{align}
The scattering terms of adjacent roots inside a string will cause the divergences
\begin{equation}
o_{a,a+1}=(\delta_{a+1}-\delta_a)^{-1}+O(1).
\end{equation}

We will consider the general case, where we perform the reduction of a single $n$-string, while keeping the other string deviations finite. The first step is to add the first $n-1$ rows to the $n^{\text{th}}$ row and then add the first $n-1$ columns of the resulting matrix to the $n^{\text{th}}$ column. The internal scattering terms within the string will cancel by performing the additions and therefore the divergent $o_{a,a+1}$ terms are not present in the $n^{\text{th}}$ row and column. The determinant will not change under this addition. The  $n^{\text{th}}$-row and column are now given by
\begin{equation}
\Phi_{an}=\Phi_{na}=
\begin{cases}
 d_a - \sum_{k=n+1}^M o_{ak}   & \text{for } a<n\\
 \sum_{l=1}^n \left[ d_l - \sum_{k=n+1}^M o_{lk} \right] & \text{for } a=n\\
 \sum_{l=1}^n o_{al}   &\text{for } a>n.
\end{cases}
\end{equation}
The first $(n-1)$ x $(n-1)$ block of the Gaudin matrix up to leading order only contains the divergent self-scattering terms
\begin{equation}
\det \Phi_0^s = \prod_a^{n-1} (\delta_{a+1}-\delta_a)^{-1}.
\end{equation}
The reduced Gaudin determinant is given by the remaining entries after cutting off the first $n-1$ rows and columns
\begin{equation}
\Phi^\text{r}_{ab} =
\begin{pmatrix}
 \sum_{l=1}^n \left[ d_l - \sum_{k=n+1}^M o_{lk} \right] &  \sum_{l=1}^n o_{lb} \\
\sum_{l=1}^n o_{al} & \tilde \Phi_{ab}
\end{pmatrix}.
\end{equation}
Using the same logic as in the derivation of the Bethe-Takahashi equations in section~\ref{sec:two},
\begin{equation}
 \sum_{l=1}^n \left[ d_l - \sum_{k=n+1}^M o_{lk} \right] = N \partial_{\alpha_n}(\theta_{n-1}(\alpha_n)+\theta_{n+1}(\alpha_n))-\sum_{l=1}^n \sum_{k=n+1}^M \partial_{\lambda_l} \theta_2(\lambda_l-\lambda_k)
\end{equation}
where $\alpha_n$ is the root center of the string and $\lambda_j$ are taken as the rapidities inside the $n$-string with zero deviations. This situation assumed the presence of only one reduced string with exponentially small deviations, but it can easily be extended to multiple reduced strings. The scattering terms between two reduced strings must then be replaced by their Bethe-Takahashi equivalents $\Theta_{nm}(\alpha^n_j-\alpha^m_k)$. The remaining rapidities outside the reduced $n$-string can still be entered in this expression including their string deviations. The divergences of the $n$-string with exponentially vanishing deviations are now extracted from the determinant using the explained reduction,
\begin{equation}
\det \Phi = \det \Phi^\text{r}  \prod_a^{n-1} (\delta_{a+1}-\delta_a)^{-1}.
\end{equation}
The divergent product will cancel against the divergences in the products present in the prefactors of the matrix elements. 

The $H$-determinant becomes indeterminate, as columns become equal to leading order as deviations get exponentially close to zero. A similar rearrangement must be applied to these determinants as well:
\begin{align}
H_{ab}(\{\mu\},\{\lambda\})&=\frac{1}{\phi_0(\mu_a-\lambda_b)}\left[ \prod_{j\neq a}\phi_2(\mu_j-\lambda_b)-\left[\frac{\phi_{-2}(\lambda_b)}{\phi_{2}(\lambda_b)}\right]^N\prod_{j\neq a}\phi_{-2}(\mu_j-\lambda_b)\right] \label{eq:HdetBethe}\\
&=\frac{1}{\phi_0(\mu_a-\lambda_b)}\left[ \prod_{j\neq a}\phi_2(\mu_j-\lambda_b)+\prod_k \frac{\phi_{-2}(\lambda_b-\lambda_k)}{\phi_{2}(\lambda_b-\lambda_k)} \prod_{j\neq a}\phi_{-2}(\mu_j-\lambda_b)\right]\\
&=K_{ab}^+ G_b^+ + K_{ab}^- G_b^- \frac{\tilde F_b^-}{\tilde F_b^+}
\end{align}
where
\begin{align}
K^\pm_{ab}&=\phi_0^{-1}(\mu_a-\lambda_b)\phi_{\pm 2}^{-1}(\mu_a-\lambda_b)\\
G_b^c&=\prod_k \phi_{2c} (\mu_k-\lambda_b)\\
\tilde F_b^c &= \prod_k \phi_{2c} (\lambda_b-\lambda_k)\\
F_b^c &= \prod_{k\neq b-c} \phi_{2c} (\lambda_b-\lambda_k).
\end{align}
By substituting the Bethe equations in equation~(\ref{eq:HdetBethe}), the expression for H is now identical to that in the spin 1/2 case. The derivation of the reduced H-matrix goes along exactly the same lines~\cite{2005_Caux_JSTAT_P09003},
\begin{align}
H^r_{ab}&=K_{ab}^- \quad \text{if } b<n \\
H^r_{an}&=\left[ \frac{F_0^0 F_1^1}{G_n^0}\prod_{j=2}^n G_j^0\right]\frac{G_b^0 G_{b+1}^0}{F_b^0 F_{b+1}^0}\left[(\delta_{b0}+\delta_{bn}-1)L_{ab}+(\delta_{b0}+\delta_{bn})K^-_{ab}\right]
\end{align}
where $L_{ab}=\partial_\mu K^-_{ab}$.
The other columns of the H-matrix remain unchanged for strings with finite deviations. The rapidities of the reduced strings are simply the rapidities of the non-deviated string. In the presence of a single reduced $n$-string:
\begin{equation}
H^r_{ab}=
\begin{cases}
K_{ab}^- & \text{if } b<n\\
H_{an}^r & \text{if } b=n\\
H_{ab} & \text{otherwise}.
\end{cases}
\end{equation}
For $H^-$-matrix for spin-1, the last column remains unchanged as well,
\begin{equation}
H^{-,r}_{ab}=
\begin{cases}
H^r_{ab} & \text{if } b<M\\
2 \phi_1^{-1}(\mu_a)\phi_{-1}^{-1}(\mu_a) & \text{if } b=M.
\end{cases}
\end{equation}

The reduction of single three-strings while keeping the remaining two-string deviations finite is to be applied to the cases where the deviations become close to the divergences in the scattering terms. In the corresponding matrix elements, the determinants should be replaced by their reduced versions, and the divergent prefactors arising from scattering terms between adjacent roots inside a reduced string should be left out. 

Finally, for finite field it is important to consider the contribution of matrix elements of lower-weight states. Highest-weight states are the Bethe states with only finite rapidities, the total spin raising operator $S^+_0$ then annihilating the state. Lower-weight states can be constructed by acting with the total spin-lowering operator $S^-_0$ on a highest weight state, in other words by adding an infinite rapidity, $| \{\lambda,\infty\}\rangle \propto S_0^- | \{\lambda\}\rangle$. The matrix elements of the corresponding lower weight states are~\cite{1981_Mueller_PRB_24}
\begin{equation}
F_q^a(\{\mu\},\{\lambda,\infty\})=\frac{\langle \{ \mu\} | [S_q^a,S_0^-] | \{\lambda\} \rangle}{\sqrt{\langle \{ \mu\} | \{\mu\} \rangle \langle \{ \lambda\} | S_0^+S_0^- | \{\lambda\} \rangle}}.
\end{equation}
For a highest-weight state, $\langle S_0^+ S_0^- \rangle=2\langle S^z_{\text{av}} \rangle$. The average magnetisation for the spin-$1$ chain is $\langle S^z_{\text{av}} \rangle = \frac{1}{N}(N-M)$. The commutators are equal to $[S_q^+,S^-_0]=\frac{2}{\sqrt{N}} S_q^z$, $[S_q^z,S^-_0]=-\frac{1}{\sqrt{N}} S_q^-$ and $[S_q^-,S^-_0]=0$.
Concluding, for a lower weight state with one infinite rapidity, the following relations hold
\begin{align}
| F_q^+ (\{\mu\}_M,\{\lambda,\infty\}_{M+1})|^2 &= \frac{2 | F_q^z (\{\mu\},\{\lambda\})|^2}{N-M},  \label{eq:lowerweightSpm} \\
| F_q^z (\{\mu\}_M,\{\lambda,\infty\}_{M})|^2 &= \frac{| F_q^- (\{\mu\},\{\lambda\})|^2}{2(N-(M-1))}, \label{eq:lowerweightSzz}\\
| F_q^- (\{\mu\}_M,\{\lambda,\infty\}_{M-1})|^2 &= 0.
\end{align}
For a lower weight state with two infinite rapidities, the norm becomes $\langle \{ \lambda\} | S_0^+ S_0^+ S_0^- S_0^- | \{\lambda\} \rangle$.
Commuting $S_0^+$ through twice and realising that we act on a highest weight state gives
\begin{align}
\langle \{ \lambda\} | S_0^+ S_0^+ S_0^- S_0^- | \{\lambda\} \rangle &=\langle \{ \lambda\} |( 8S^z_\text{av}S^z_\text{av}-\frac{4}{N}S^z_\text{av} )| \{\lambda\} \rangle\nonumber\\
&= \frac{8}{N^2}\left(N-M_\lambda\right)\left(N-M_\lambda-\frac{1}{2}\right) \langle \{ \lambda\} | \{ \lambda\} \rangle.
\end{align}
The only nonvanishing matrix element for a state containing two infinite rapidities is $F^+$, where the corresponding highest weight state is in the subsector of $F^-$, such that $M_\lambda=M_\mu-1$.
\begin{align}
| F_q^+ (\{\mu\}_M,\{\lambda,\infty,\infty\}_{M+1})|^2 &= \frac{\frac{4}{N^2} | F_q^- (\{\mu\},\{\lambda\})|^2}{\frac{8}{N^2}\left(N-M_\lambda\right)\left(N-M_\lambda-\frac{1}{2}\right)} \nonumber \\
&= \frac{ | F_q^- (\{\mu\},\{\lambda\})|^2}{2\left(N-M_\mu+1\right)\left(N-M_\mu+\frac{1}{2}\right)} \label{eq:lowerweighttworap}
\end{align}

The results imply that the $S^{-+}$ structure factor only contains contributions from highest-weight states,
while the $S^{zz}$ structure factor contains contributions from highest-weight states and states containing one infinite rapidity,
and the $S^{+-}$ structure factor contains contributions from highest-weight states, states containing one infinite rapidity and states containing two infinite rapidities.

\section{Dynamical structure factor}
\label{sec:results}
The results of the main computation on the dynamical structure factor for the Babujan-Takhtajan spin-$1$ chain defined in equation~(\ref{eq:defDSF1}) will be given in this section. We restrict to the case of zero temperature. The sum over the matrix elements runs only over a selected part of the Hilbert space, containing specifically all two-spinon and four-spinon contributions. The ground state is excluded from the sum as we take the connected correlation function. 

The sum of all included matrix element contributions irrespective of the their energy and momentum provides a quantitative measure of the quality of the computed dynamical structure factors. A comparison with an analytic expression for the integrated density of the dynamical structure factor in equation~(\ref{eq:defDSF1}) yields a saturation value for the sum rule of the corresponding computation,
\begin{equation}
t^{a \bar a} \equiv \int_{-\infty}^{\infty} \frac{\text{d}\omega}{2\pi}\frac{1}{N}\sum_q S^{a \bar a}(q,\omega)=\langle S^aS^{\bar a}\rangle_c.
\end{equation}
For the various structure factors the integrated densities are $t^{\pm \mp}=\langle S^\pm S^\mp \rangle$ and $t^{zz}=\langle S^zS^z\rangle-\langle S^z\rangle^2$. In the spin-$1$ representation of local spin matrices $S^-S^+=2-S^z S^z-S^z$ holds, from which follows
\begin{equation}
t^{\pm \mp}=2-t^{zz} -\langle S^z \rangle^2 \pm\langle S^z \rangle. \label{eq:tpmtzzsumrule}
\end{equation}

Unfortunately, it is not possible to derive sum rules for each structure factor independently from each other in finite field. However, for zero field simplifications can be made. Due to the spin rotational symmetry, the structure factors in the transverse and longitudinal direction become equal up to a factor of two, $t^{zz}=\frac{1}{2}t^{-+}$, such that
\begin{equation}
t^{-+}=t^{+-}=\frac{4}{3}, \qquad
t^{zz}=\frac{2}{3}.
\label{eq:sumruledef}
\end{equation}

When including all states including all string deviations, the sum of all matrix elements must be exactly equal to sum rule~(\ref{eq:sumruledef}). We solve the Bethe states up to arbitrary high precision including all string deviations, while determinants of the matrix elements will be computed to the same high precision as well. Herefore we make use of the arbitrary precision computation library ARPREC~\cite{ARPREC}. The saturation for the matrix element contributions of several types of Bethe states have been computed for small system sizes, for zero field as well as finite field. For the latter, matrix element contributions for both the longitudinal and transverse dynamical structure factor have been computed, while matrix element contributions from lower weight states must be included as well. In both cases, we saturate the available sum rules to exactly $100\%$ up to arbitrary precision making use of the ARPREC algorithm. In appendix~\ref{app:arprecA} we state our sum rule saturation results up to $64$ digits for $4$ and $6$ lattice sites. Without the correct treatment of string deviations, this exact saturation would not have been possible. The exact sum rule saturation of our computations completes a consistency check on the algorithm.

Furthermore, we provide numerical evidence for the statement that matrix elements of singular pair states vanish~\cite{2007_Hagemans_JPA_40}. We saturate the sum rule to identically one, where all matrix elements of singular pair states were not taken into account, implying that their contributions must vanish identically. By the same token, we can imply that matrix elements of exceptional solutions to the Bethe equations for spin-$1$ vanish as well.

\begin{figure}[t!]
\centering
\includegraphics{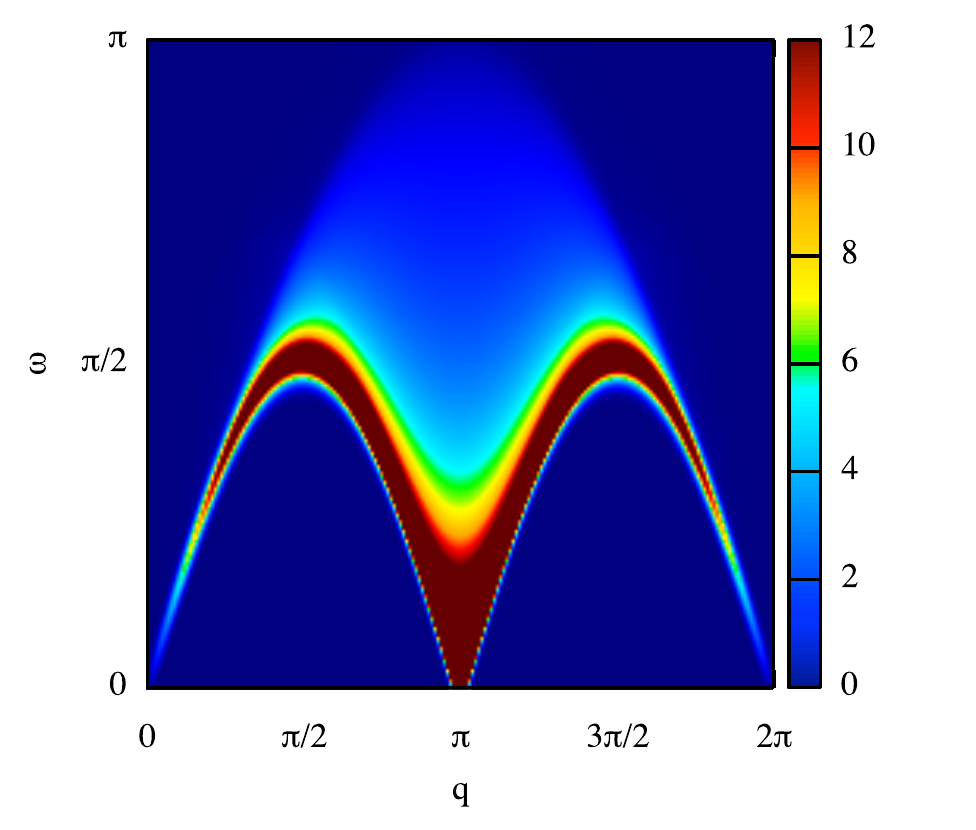}
\caption{Transverse dynamical structure factor of the Babujan-Takhtajan spin-$1$ chain at $N=200$, including two-spinon and four-spinon contributions, with a sum rule saturation of $99.16\%$.}
\label{fig:twospinon}
\end{figure}

We continue by computing the dynamical structure factor for larger system sizes. We restrict to zero field, where only the transverse dynamical structure factor is relevant. It is not possible to include all states anymore as the Hilbert space becomes exponentially large, so we restrict to the spinon states described in table~\ref{table:extrans}. For exponentially small deviations close to the divergencies in the matrix element expressions, reduced expressions must be used. Only if the three-string deviations become smaller than $O(10^{-8})$, we put the deviation to zero and use the reduced formalism. The algorithm keeps track of the remaining two-string deviations at all times. 

To be able to represent the dynamical structure factor in graphics, the delta functions need to be smoothened by Gaussians,
\begin{equation}
\delta(\omega-\omega_\alpha)=\frac{1}{\sqrt{\pi}\epsilon}e^{-(\omega-\omega_\alpha)^2/\epsilon^2}
\end{equation}
where the value of the width $\epsilon$ is of order $1/N$. Figure~\ref{fig:twospinon} shows the transverse dynamical structure factor for a system size of 200 sites. At this system size, the sum rule contribution of the two-spinon states containing two-strings and a single real rapidity is $89.88\%$.

\begin{figure}[t!]
\centering
\includegraphics{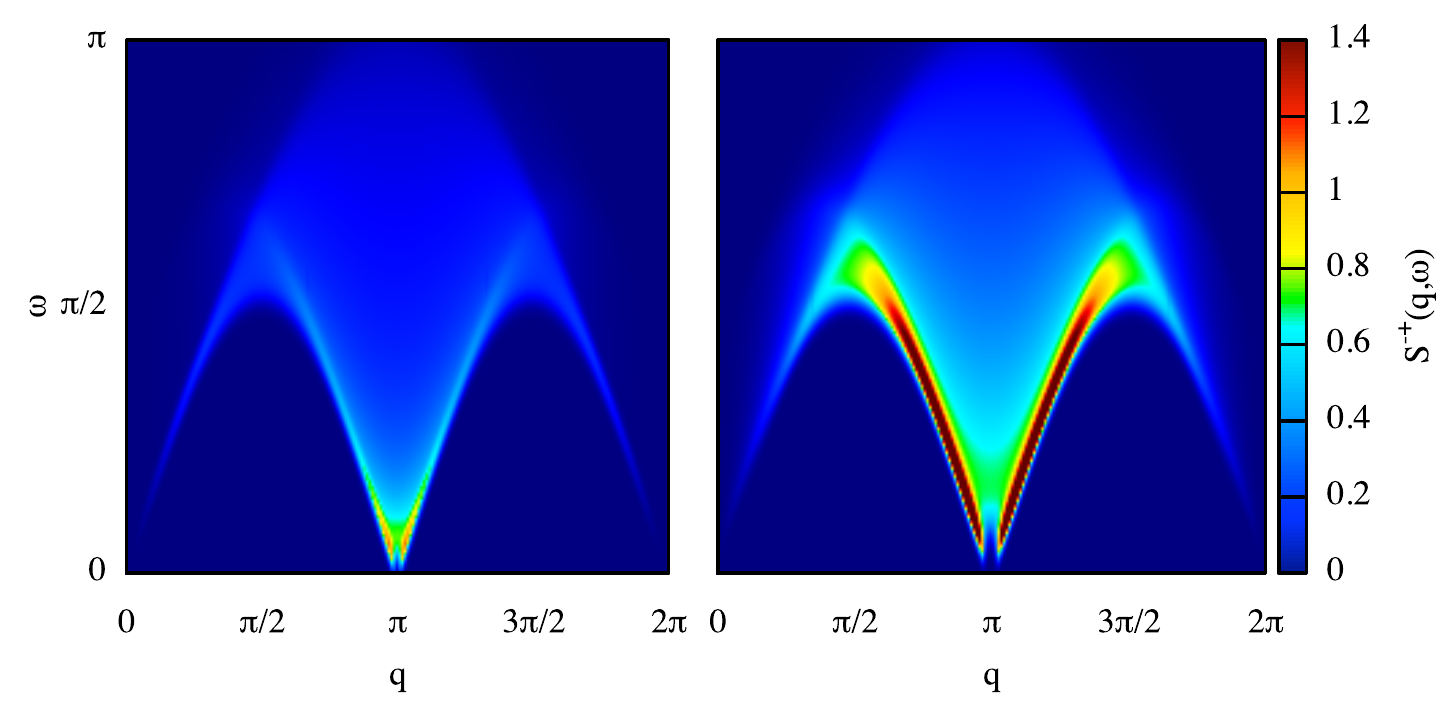}
\caption{Four-spinon contributions to the transverse dynamical structure factor at $N=200$. Left: two-strings and a single three-string. The sum rule contribution is $2.80\%$. Right: two-strings, a single three-string and two one-strings (real rapidities). The sum rule contribution is $6.47\%$.}
\label{fig:fourspinon}
\end{figure}

The two different types of four-spinon contributions are shown separately in figure~\ref{fig:fourspinon} and yield a sum rule contribution of $2.80\%$ and $6.47\%$ respectively. The total sum rule saturation of the computation is $99.16\%$.  

\begin{figure}[t!]
\centering
\includegraphics{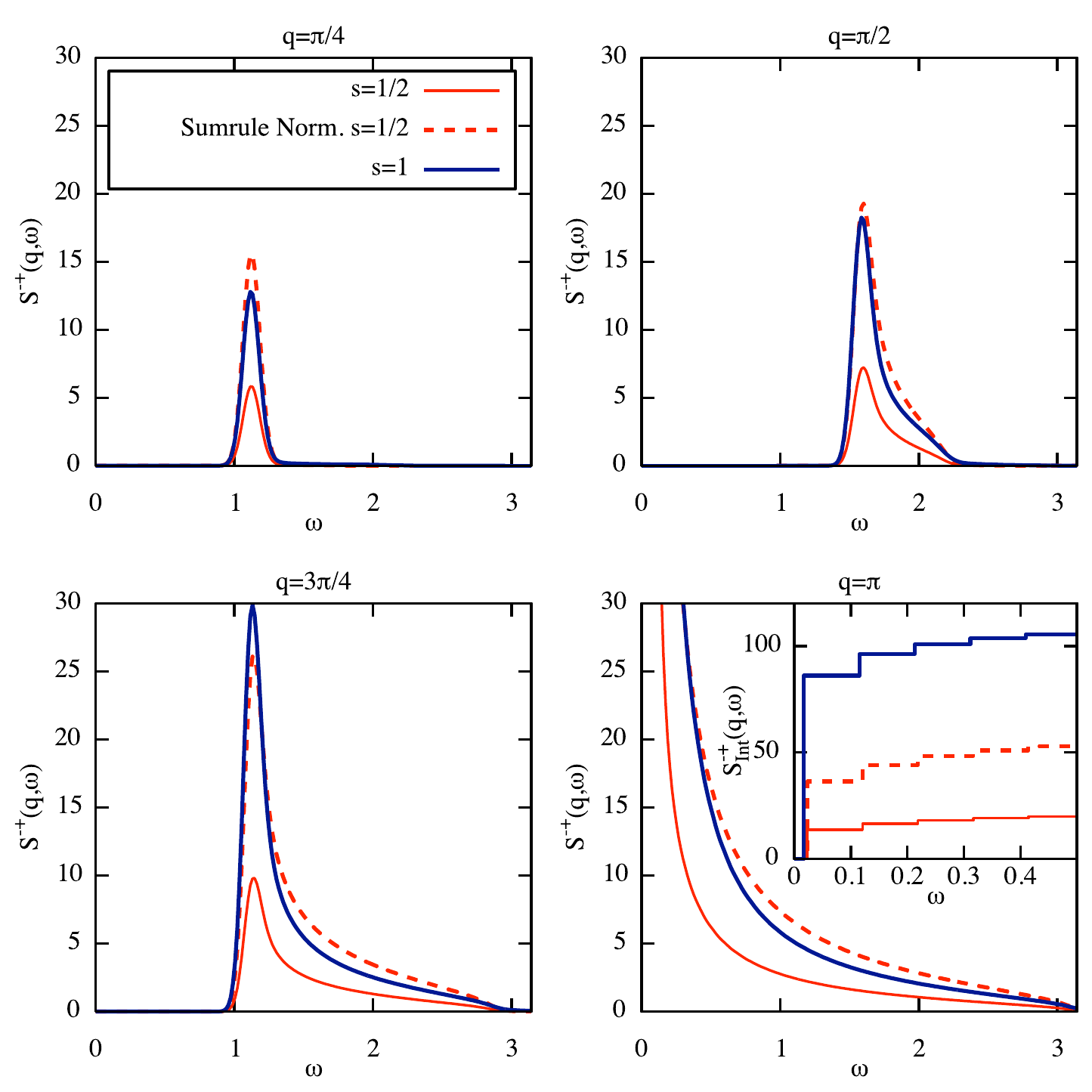}
\caption{Fixed momentum cuts of the transverse dynamical structure factor at $N=200$ for spin-1 (sum rule saturation $99.16\%$), spin-$\frac{1}{2}$ (sum rule saturation $99.24\%$) and spin-$\frac{1}{2}$ normalised to the spin-$1$ sum rule. Inset: $S_\text{Int}=\int_0^\omega \text{d}\omega^\prime S(k,\omega^\prime)=2\pi\sum_{\omega_\alpha < \omega} | F_\alpha |^2$.}
\label{fig:fixedP}
\end{figure}

Figure~\ref{fig:fixedP} shows fixed momentum cuts of the dynamical structure factor, where the data is being compared to ABACUS~\cite{2005_Caux_JSTAT_P09003,2005_Caux_PRL_95} results for the spin-$\frac{1}{2}$ case. Due to a lower integrated density, it is obvious that the spin-$\frac{1}{2}$ dynamical structure factor is smaller than the spin-$1$ data at all momenta and energies. Therefore, to be able to make a comparison of the shape of the correlations, we normalise the spin-$\frac{1}{2}$ dynamical structure factor to the spin-$1$ sum rule. At small momenta, the normalised spin-$\frac{1}{2}$ correlations are above the spin-$1$ data, while with increased momenta towards $q=\pi$, the spin-$1$ correlations are higher near the lower boundaries of the spectrum. The correlation remains below the normalised spin-$\frac{1}{2}$ correlations at higher energies.

At $q=\pi$, we show a cumulative plot of the matrix elements for states with the smallest energies in figure~\ref{fig:fixedP}. Again, the spin-$1$ dynamical structure factor is higher close to the lower boundary, while it lies below the normalised spin-$\frac{1}{2}$ dynamical structure factor at higher energies.

\begin{figure}[h!]
\centering
\includegraphics{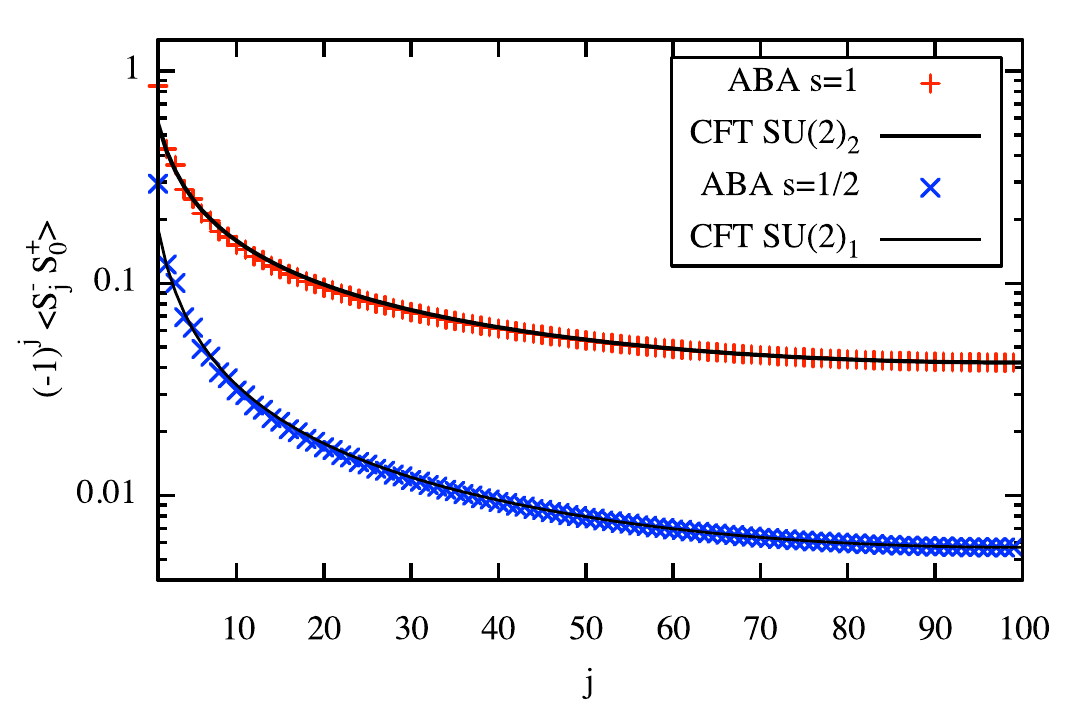}
\caption{Equal time spin-spin correlation at $N=200$ with sum rule saturation $99.16\%$ for spin-$1$, compared to the prediction from the SU$(2)$ level $2$ WZW-model. The prefactor has been used as fitting parameter only. The same comparison is shown for spin-$\frac{1}{2}$ data from ABACUS at $N=200$ (sum rule saturation $99.24\%$) and the SU$(2)$ level $1$ WZW-model.}
\label{fig:cft}
\end{figure}

The real space-time dependent correlation function can be obtained by inverse Fourier transforming the results,
\begin{equation}
\langle S^a_j(t) S^{\bar a}_0(0) \rangle = \frac{1}{N} \sum_\alpha |F^a_{q_\alpha}|^2 e^{-i q_\alpha j - i \omega_\alpha t}
\end{equation}
and is plotted in figure~\ref{fig:cft}. We compare our results at equal time to the predictions on the asymptotics from the continuum limit described by conformal field theory. The integrable chains of spin-$s$ have a critical low-energy sector \cite{1986_Affleck_NPB_265,1987_Solyom_PRB_36,1988_Affleck_CMP_115,1986_Affleck_PRL_56_1,1987_Affleck_PRB_36,1988_Alcaraz_JPA_21,1990_Tsvelik_PRB_42_16,2013_Michaud_PRB_87} described by the SU(2) level $2s$ Wess-Zumino-Novikov-Witten models \cite{1984_Knizhnik_NPB_247}. Under non-Abelian bosonization, the spin-spin correlations are asymptotically given by those of the fundamental WZNW primary fields, this leading to the prediction that in an infinite system, the dominant antiferromagnetic correlations decay as a power law
\begin{equation}
\langle S^a_j S^{\bar{a}}_{0} \rangle \sim \frac{(-1)^j}{|j|^{2\Delta}},
\end{equation}
in which the scaling dimension $\Delta=h+\bar h$ can be obtained from the primary field scaling dimension for a general SU($n$) level k WZW model \cite{1984_Knizhnik_NPB_247},
\begin{equation}
h=\bar h = \frac{n^2-1}{2n(n+k)}.
\end{equation}
Substituting the values for level $k=2$ and performing a conformal mapping to finite size so that the distance function becomes $j \rightarrow \frac{N}{\pi}\sin(j \frac{\pi}{N})$, the conformal field theory prediction for the asymptotics of the correlations of the Babujan-Takhtajan chain is thus
\begin{equation}
\langle S^a_j S^{\bar a}_0 \rangle \sim {(-1)}^j \left[\frac{N}{\pi}\sin\left(j \frac{\pi}{N}\right) \right]^{-\frac{3}{4}},
\end{equation}
up to a non-universal prefactor and subleading corrections. 
We plot our results against this prediction in figure \ref{fig:cft}, where the pre factor has been used as the only fitting parameter. For comparison, we also give the corresponding fits for the spin-$1/2$ case. As is clearly seen, the agreement is excellent over all distances but the very smallest. 

\section{Conclusions}
The dynamical structure factor of the Babujan-Takhtajan spin-1 chain has been computed numerically at zero field and zero temperature. Correct treatment of the string deviations makes it possible to obtain the roots of the Bethe equations up to high precision. Two-spinon and four-spinon Bethe states have been constructed by perturbing the sea of two-strings from the ground state with one-strings and three-string respectively. The matrix elements of these states are shown to provide over $99\%$ of all contributions to the dynamical structure factor at $N=200$. 

Our work for the dynamical structure factor of the spin-$1$ Babujan-Takhtajan chain could be extended to finite magnetic field, by summing over the matrix elements of the important contributing Bethe states.  Another interesting possibility would be to compute two-spinon contributions in the thermodynamic limit directly from the isotropic limit of results from the vertex operator approach \cite{1994_Bougourzi_NPB_417,1994_Konno_NPB_432,JimboBOOK}. The anisotropic case of the spin-$1$ chain~\cite{FateevZamol,1990_Frahm_NPB_336} might provide avenues for further extensions. Moreover, due to the existence of matrix element expressions for chains of arbitrary spins at each lattice site, generalisation of our method to mixed, alternating or impurity spin chains would be within reach as well. 

\section*{Acknowledgements}
We thank M. Brockmann, H. Konno, R. Weston and J. Willetts for useful discussions. 
The authors acknowledge support from the Dutch Foundation for Fundamental Research on Matter (FOM) and from the Netherlands Organisation for Scientific Research (NWO). We thank SURFsara for the support in using the Lisa Compute Cluster for our computations.

\bibliographystyle{unsrt}
\addcontentsline{toc}{chapter}{Bibliography}

\bibliography{}

\newpage

\appendix
\section{Sum rule saturation results}
\label{app:arprecA}
The sum rule of the matrix elements of different types of Bethe states is calculated by $t^{a \bar a}= \frac{1}{N} \sum_{\alpha} |F^a_q|^2$. 

One exceptional solution to the Bethe equations is within the states containing one one-string and one three-string in table~\ref{table:N6M4sat}, where for the symmetric state the two real rapidities coincide. The saturation of the sum rule of the remaining states to identically $100\%$ indicates that the matrix element of the exceptional state must vanish.
\begin{table}[h!]
\scriptsize
\centering
\begin{tabular}{l|l}
State & $t^{-+}$\\[0.3em]
\toprule
 \parbox{0.2\textwidth}{
\begin{itemize}
\item 1 one-string
\item 1 two-string
\end{itemize} }& $1.3298902414257309778968260082398925737462210732701764927080799842$\\
\midrule
 \parbox{0.2\textwidth}{
\begin{itemize}
\item 1 three-string
\end{itemize} }& $0.0034430919076023554365073250934407595871122600631568406252533493$\\
 \bottomrule
 \end{tabular}
 \captionsetup{width=.65\textwidth}
\caption{$N=4$ zero field transverse sum rule saturation of all Bethe states. $t^{-+}=\frac{4}{3}$.}
\end{table}

\begin{table}[h!]
\scriptsize
\centering
\begin{tabular}{l|l}
State & $t^{-+}$\\[0.3em]
\toprule
 \parbox{0.2\textwidth}{
\begin{itemize}
\item 1 one-string
\item 2 two-strings
\end{itemize} }& $1.3255257488748488860163283692758424230579681706120362031941046259$\\
\midrule
 \parbox{0.2\textwidth}{
\begin{itemize}
\item 1 two-string
\item 1 three-string
\end{itemize} }& $0.0075055971882272731525334425805834021031872300622757573898167819$\\
\midrule
 \parbox{0.2\textwidth}{
\begin{itemize}
\item 2 one-strings
\item 1 three-string
\end{itemize} }& $0.0002622186877160566127125335739058592402906351705930607556314381$\\
\midrule
 \parbox{0.2\textwidth}{
\begin{itemize}
\item 1 one-string
\item 1 four-string
\end{itemize} }& $0.0000397543061725485159834550328504808640766215971193428953405972$\\
\midrule
 \parbox{0.2\textwidth}{
\begin{itemize}
\item 1 five-string
\end{itemize} }& $0.0000000142763685690357755328701511680678106758913089690984398903$\\
 \bottomrule
 \end{tabular}
 \captionsetup{width=.65\textwidth}
\caption{$N=6$ zero field transverse sum rule saturation of all Bethe states. $t^{-+}=\frac{4}{3}$.}
\end{table}

\begin{table}[h!]
\scriptsize
\centering
\begin{tabular}{l|l}
State & $t^{-+}$\\[0.3em]
\toprule
 \parbox{0.2\textwidth}{
\begin{itemize}
\item 1 one-string
\end{itemize} }& $0.6846990312590646393713079482707358515307377231873608519159514473$\\[0.5em]
State & $t^{zz}_{HW}$\\[0.3em]
\toprule
 \parbox{0.2\textwidth}{
\begin{itemize}
\item 2 one-strings
\end{itemize} }& $0$\\
\midrule
 \parbox{0.2\textwidth}{
\begin{itemize}
\item 1 two-string
\end{itemize} }& $0.4511844635310912540668073936841415065474726562814123394313899782$\\
 \bottomrule
 \end{tabular}
 \captionsetup{width=.65\textwidth}
\caption{$N=4,M=2$ finite field sum rule saturation. The contribution of lower weight states is incorporated in the sum rule. $\quad t^{zz}_\text{HW}=\frac{5}{4}-\frac{7}{6}t^{-+}$.}
\end{table}

\begin{table}[h!]
\scriptsize
\centering
\begin{tabular}{l|l}
State & $t^{-+}$\\[0.3em]
\toprule
 \parbox{0.2\textwidth}{
\begin{itemize}
\item 1 one-string
\item 1 two-string
\end{itemize} }& $0.9114222449196104572254525102473452066445024764065529440308002752$\\[0.5em]
\midrule
 \parbox{0.2\textwidth}{
\begin{itemize}
\item 3 one-strings
\end{itemize} }& $0.0004074719549408679440733606676982892360610948842848397581256742$\\[0.5em]
\midrule
 \parbox{0.2\textwidth}{
\begin{itemize}
\item 1 three-string
\end{itemize} }& $0.0002114163546684467910556012520301073762100126243669746431148786$\\[0.5em]
State & $t^{zz}_{HW}$\\[0.3em]
\toprule
 \parbox{0.2\textwidth}{
\begin{itemize}
\item 2 one-strings
\item 1 two-string
\end{itemize} }& $0.0002848110766734939157393853106721216300909654070606992945276071$\\
\midrule
 \parbox{0.2\textwidth}{
\begin{itemize}
\item 2 two-strings
\end{itemize} }& $0.4814244133454303130164163967970538203043691866313212111429643782$\\
\midrule
\parbox{0.2\textwidth}{
\begin{itemize}
\item 1 one-string
\item 1 three-string
\end{itemize} }& $0.0097493792684611458741399625435615923330214237710012341463743817$\\
\midrule
\parbox{0.2\textwidth}{
\begin{itemize}
\item 1 four-string
\end{itemize} }& $0.0000489630975675354619147600426821508218381318451001928009748890$\\
 \bottomrule
 \end{tabular}
 \captionsetup{width=.65\textwidth}
\caption{$N=6,M=4$ finite field sum rule saturation. The contribution of lower weight states is incorporated in the sum rule. $\quad t^{zz}_\text{HW}=\frac{14}{9}-\frac{7}{6}t^{-+}$.
}
\label{table:N6M4sat}
\end{table}

\newpage
$\phantom{A}$
\newpage
\section{Tables of rapidities and matrix elements}
\label{app:tables}
This section provides tables of rapidities and their corresponding matrix elements of the dynamical structure factor in both zero field and finite field for small system sizes consisting of $4$ and $6$ sites respectively. For $|F^{a \bar a}_M|^2$, $M$ denotes in which spin sub sector the ground state is taken as a reference state for the matrix elements. The momentum of the Bethe states is given by
\begin{align}
q &= M\pi + \frac{2\pi}{N} \sum_{j=1}^{M} J_j \mod 2\pi\\ 
P &=\frac{N}{2\pi}q =M\frac{N}{2}+\sum_{j=1}^{M} J_j \mod N.
\end{align}
\begin{table}[h!]
\scriptsize
\centering
\begin{tabular}{rrrrrr}
BT & B & $\lambda$& E & P & $|F^{-+}_{M2}|^2$ \\[0.3em]
\toprule
-1 & -1 & $-1.$& $-0.5$ & 1 & $0.1081941875543878$  \\
\midrule
0 & 0 & $0.$& $-1.$ & 2 & $2.5224077499274829$  \\
\midrule
1 & 1 & $1.$& $-0.5$ & 3 & $0.1081941875543878$  \\
 \bottomrule
 \end{tabular}
 \captionsetup{width=.65\textwidth}
\caption{$N=4$, $M=1$, 3 states}
\end{table}
\begin{table}[h!]
\scriptsize
\centering
\begin{tabular}{rrrrrr}
BT & B & $\lambda$& E & P & $|F^{zz}_{M2}|^2$ \\[0.3em]
\toprule
$-2_2$ & $-1.5$ & $0.8164965809277260+0.5773502691896258i$ & $-1$ & $2$ & $0.1321488698022421$ \\
 & $-0.5$ & $0.8164965809277260-0.5773502691896258i$ &  &  &  \\
\midrule
$-1_2$ & $-0.5$ & $-0.3333333333333333+0.4714045207910317i$ & $-2$ & $3$ & $0.7702200572599404$ \\
 & $-0.5$ & $-0.3333333333333333-0.4714045207910317i$ &  &  &  \\
\midrule
$0_2$ & $-0.5$ & $0.5110810845293939i$ & $-2.7071067811865475$ & $0$ & - \\
 & $0.5$ & $-0.5110810845293939i$ &  &  &  \\
\midrule
$1_2$ & $0.5$ & $0.3333333333333333+0.4714045207910317i$ & $-2$ & $1$ & $0.7702200572599404$ \\
 & $0.5$ & $0.3333333333333333-0.4714045207910317i$ &  &  &  \\
\midrule
$2_2$ & $0.5$ & $-0.8164965809277260+0.5773502691896258i$ & $-1$ & $2$ & $0.1321488698022421$ \\
 & $1.5$ & $-0.8164965809277260-0.5773502691896258i$ &  &  &  \\
 \midrule
$0.5_1$ & $0.5$ & $0.7395391542562349$ & $-1.2928932188134525$ & $0$ & $0$ \\
$-0.5_1$ & $-0.5$ & $-0.7395391542562349$ &  &  &  \\
  \bottomrule
 \end{tabular}
 \captionsetup{width=.65\textwidth}
\caption{$N=4$, $M=2$, 6 states}
\end{table}
\begin{table}[h!]
\scriptsize
\centering
\begin{tabular}{rrrrrrr}
BT & B & $\lambda$& E & P & $|F^{+-}_{M2}|^2$ & $|F^{-+}_{M4}|^2$ \\[0.3em]
\toprule
$0_1$ & $0$ & $-0.4597842760651707$& $-2.3090169943749474$ & 3 & $0.0032938977617409$ & 0.7747958321493609 \\
$1_2$ & $0$ & $0.5389091324075328+0.5574169414591360i$&  &  &  &  \\
 & $1$ & $0.5389091324075328-0.5574169414591360i$&  &  &  &  \\
\midrule
$0_1$ & $0$ & $0$& $-3.5$ & 2 & $3.0899494936611665$ & 3.7699693014042021 \\
$0_2$ & $0$ & $0.4472135954999579i$&  &  &  &  \\
 & $0$ & $-0.4472135954999579i$&  &  &  &  \\
\midrule
$0_1$ & $0$ & $0.4597842760651707$& $-2.3090169943749474$ & 1 & $0.0032938977617409$ & 0.7747958321493609 \\
$-1_2$ & $-1$ & $-0.5389091324075328+0.5574169414591360i$&  &  &  &  \\
 & $0$ & $-0.5389091324075328-0.5574169414591360i$&  &  &  &  \\
\midrule
$-1_3$ & $0$ & $-0.5546486453949190$& $-1.1909830056250526$& 3 &	$0.3274672866957473$ & 0.0068861838152047 \\
 & $-2$ & $-0.5316926716774879+1.0109029892426487i$&  &  &  & \\
 & $-1$ & $-0.5316926716774879-1.0109029892426487i$& &  &  & \\
 \midrule
$0_3$ & - & $0$& $-1.5$& 0 &	$0$ & 0 \\
 & - & $i$&  &  &  & \\
 & - & $-i$& &  &  & \\
 \midrule
$1_3$ & 0 & $0.5546486453949190$& $-1.1909830056250526$& 1 &	$0.3274672866957473$ & 0.0068861838152047 \\
 & 1 & $0.5316926716774879+1.0109029892426487i$&  &  &  & \\
 & 2 & $0.5316926716774879-1.0109029892426487i$& &  &  & \\
 \bottomrule
 \end{tabular}
 \captionsetup{width=.65\textwidth}
\caption{$N=4$, $M=3$, 6 states}
\end{table}

\begin{table}[h!]
\scriptsize
\centering
\begin{tabular}{rrrrrr}
BT & B & $\lambda$& E & P & $|F^{zz}_{M4}|^2$ \\[0.3em]
\toprule
$-0.5_2$ & $-0.5$  & $-0.2958366877513515+0.5512945305070179i$ & $-4.3507810593582122$ & $0$ & $-$(GS) \\
 & $0.5$ & $-0.2958366877513515-0.5512945305070179i$ &  &  &  \\
$0.5_2$& $-0.5$ & $0.2958366877513515+0.5512945305070179i$ &  &  &  \\
 & $0.5$ & $0.2958366877513515-0.5512945305070179i$ &  &  &  \\
\midrule
$0_4$ & $-0.5$ & $0.5003904510332184i$ & $-1.1492189406417878$ & $0$ & $0$ \\
 & $0.5$ & $-0.5003904510332184i$ &  &  &  \\
 & $0.5$ & $1.5221027460129776i$ &  &  &  \\
 & $-0.5$ & $-1.5221027460129776i$ &  &  &  \\
 \midrule
$0_1$ & - & $0$ & $-2.5$ & $0$ & $0$ \\
$0_3$ & - & $0$ &  &  &  \\
 & - & $i$ &  &  &  \\
 & - & $-i$ &  &  &  \\
 \bottomrule
 \end{tabular}
 \captionsetup{width=.65\textwidth}
\caption{$N=4$, $M=4$, 3 states}
\end{table}

\begin{table}[h!]
\scriptsize
\centering
\begin{tabular}{rrrrrr}
BT & B & $\lambda$& E & P & $|F^{-+}_{M4}|^2$ \\[0.3em]
\toprule
$-1_1$ & $-1$ & $-0.2555277805066051$ & $-1.6344017903173264$ & $5$ & $0.0001789723688635$ \\
$-3_2$ & $-2$ & $-1.1236418771967856+0.6063626703700935i$ & & \\
& $-1$ & $-1.1236418771967856-0.6063626703700935i$ & & \\
\midrule
$0_1$ & $0$ & $0.3686711262545637$ & $-1.5$ & $0$ & $0.0000000000000000$ \\
$-3_2$ & $-2$ & $-1.2449957349071031+0.6066082587665470i$ & & \\
& $-1$ & $-1.2449957349071031-0.6066082587665470i$ & & \\
\midrule
$1_1$ & $1$ & $1.3499213155394399$ & $-0.9224880444272917$ & $1$ & $0.0000761620641837$ \\
$-3_2$ & $-2$ & $-1.3370590369569984+0.6110017571295619i$ & & \\
& $-1$ & $-1.3370590369569984-0.6110017571295619i$ & & \\
\midrule
$-1_1$ & $-1$ & $-0.8382025477478952$ & $-2.5$ & $0$ & $0.0000000000000000$ \\
$-2_2$ & $-1$ & $-0.3714681411681472+0.4891628925301364i$ & & \\
& $-1$ & $-0.3714681411681472-0.4891628925301364i$ & & \\
\midrule
$0_1$ & $0$ & $0.2777436124668177$ & $-2.2306475087141597$ & $1$ & $0.0076129377639565$ \\
$-2_2$ & $-1$ & $-0.6636155913814564+0.4657466394203293i$ & & \\
& $-1$ & $-0.6636155913814564-0.4657466394203293i$ & & \\
\midrule
$1_1$ & $1$ & $1.2563934702520216$ & $-1.5616187257160437$ & $2$ & $0.0025479594342444$ \\
$-2_2$ & $-1$ & $-0.7483270479200486+0.4646161617183808i$ & & \\
& $-1$ & $-0.7483270479200486-0.4646161617183808i$ & & \\
\midrule
$-1_1$ & $-1$ & $-1.0194087434039865$ & $-3.0922399830038464$ & $1$ & $0.0363666497936189$ \\
$-1_2$ & $-1$ & $-0.0951457978234211+0.5024452796149906i$ & & \\
& $0$ & $-0.0951457978234211-0.5024452796149906i$ & & \\
\midrule
$0_1$ & $0$ & $0.1738705041478744$ & $-3.1495526854710696$ & $2$ & $0.6396985716624094$ \\
$-1_2$ & $-1$ & $-0.2782612070695909+0.5060767423118732i$ & & \\
& $0$ & $-0.2782612070695909-0.5060767423118732i$ & & \\
\midrule
$1_1$ & $1$ & $1.1761413042818725$ & $-2.3660254037844386$ & $3$ & $0.0228050841164643$ \\
$-1_2$ & $-1$ & $-0.3612605080004833+0.5051495227671889i$ & & \\
& $0$ & $-0.3612605080004833-0.5051495227671889i$ & & \\
\midrule
$-1_1$ & $-1$ & $-1.1032124888323004$ & $-3$ & $2$ & $0.1630928137396166$ \\
$0_2$ & $0$ & $0.1185935425239309+0.4975975543410103i$ & & \\
& $0$ & $0.1185935425239309-0.4975975543410103i$ & & \\
\midrule
$0_1$ & $0$ & $0$ & $-3.6513878188659973$ & $3$ & $3.7237751676309478$ \\
$0_2$ & $0$ & $0.4956592188330808i$ & & \\
& $0$ & $-0.4956592188330808i$ & & \\
\midrule
$1_1$ & $1$ & $1.1032124888323004$ & $-3$ & $4$ & $0.1630928137396166$ \\
$0_2$ & $0$ & $-0.1185935425239309+0.4975975543410103i$ & & \\
& $0$ & $-0.1185935425239309-0.4975975543410103i$ & & \\
\midrule
$-1_1$ & $-1$ & $-1.1761413042818725$ & $-2.3660254037844386$ & $3$ & $0.0228050841164643$ \\
$1_2$ & $0$ & $0.3612605080004833+0.5051495227671889i$ & & \\
& $1$ & $0.3612605080004833-0.5051495227671889i$ & & \\
\midrule
$0_1$ & $0$ & $-0.1738705041478744$ & $-3.1495526854710696$ & $4$ & $0.6396985716624094$ \\
$1_2$ & $0$ & $0.2782612070695909+0.5060767423118732i$ & & \\
& $1$ & $0.2782612070695909-0.5060767423118732i$ & & \\
\midrule
$1_1$ & $1$ & $1.0194087434039865$ & $-3.0922399830038464$ & $5$ & $0.0363666497936189$ \\
$1_2$ & $0$ & $0.0951457978234211+0.5024452796149906i$ & & \\
& $1$ & $0.0951457978234211-0.5024452796149906i$ & & \\
\midrule
$-1_1$ & $-1$ & $-1.2563934702520216$ & $-1.5616187257160437$ & $4$ & $0.0025479594342444$ \\
$2_2$ & $1$ & $0.7483270479200486+0.4646161617183808i$ & & \\
& $1$ & $0.7483270479200486-0.4646161617183808i$ & & \\
\midrule
$0_1$ & $0$ & $-0.2777436124668177$ & $-2.2306475087141597$ & $5$ & $0.0076129377639565$ \\
$2_2$ & $1$ & $0.6636155913814564+0.4657466394203293i$ & & \\
& $1$ & $0.6636155913814564-0.4657466394203293i$ & & \\
\bottomrule
 \end{tabular}
 \captionsetup{width=.7\textwidth}
\caption{$N=6$, $M=3$, part I, 36 states}
\end{table}

\begin{table}[h!]
\scriptsize
\centering
\begin{tabular}{rrrrrr}
BT & B & $\lambda$& E & P & $|F^{zz}_{M4}|^2$ \\[0.3em]
\toprule
$1_1$ & $1$ & $0.8382025477478952$ & $-2.5$ & $0$ & $0.0000000000000000$ \\
$2_2$ & $1$ & $0.3714681411681472+0.4891628925301364i$ & & \\
& $1$ & $0.3714681411681472-0.4891628925301364i$ & & \\
\midrule
$-1_1$ & $-1$ & $-1.3499213155394399$ & $-0.9224880444272917$ & $5$ & $0.0000761620641837$ \\
$3_2$ & $1$ & $1.3370590369569984+0.6110017571295619i$ & & \\
& $2$ & $1.3370590369569984-0.6110017571295619i$ & & \\
\midrule
$0_1$ & $0$ & $-0.3686711262545637$ & $-1.5$ & $0$ & $0.0000000000000000$ \\
$3_2$ & $1$ & $1.2449957349071031+0.6066082587665470i$ & & \\
& $2$ & $1.2449957349071031-0.6066082587665470i$ & & \\
\midrule
$1_1$ & $1$ & $0.2555277805066051$ & $-1.6344017903173264$ & $1$ & $0.0001789723688635$ \\
$3_2$ & $1$ & $1.1236418771967856+0.6063626703700935i$ & & \\
& $2$ & $1.1236418771967856-0.6063626703700935i$ & & \\
\midrule
$-1_1$ & $-1$ & $-1.1648129344771764$& $-1.8486121811340027$ & 3 & $0.0024448317296452$ \\
$0_1$ & $0$ & $0$&  &  & \\
$1_1$ & $1$ & $1.1648129344771764$&  &  &  \\
\midrule
$-3_3$ & $-1$ & $-1.3751320020322765$ & $-0.6339745962155614$ & $3$ & $0.0000663240402261$ \\
& $-3$ & $-1.2995316710963794+1.0960218400182442i$ & & \\
& $-2$ & $-1.2995316710963794-1.0960218400182442i$ & & \\
\midrule
$-2_3$ & $-2$ & $-0.7780251893799998$ & $-1.0388285888128867$ & $4$ & $0.0005223047734948$ \\
& $-2$ & $-0.7884690768841303+1.0095363580274479i$ & & \\
& $-1$ & $-0.7884690768841303-1.0095363580274479i$ & & \\
\midrule
$-1_3$ & $-1$ & $-0.3600822571547615$ & $-1.3702226735373758$ & $5$ & $0.0000456202502845$ \\
& $-2$ & $-0.3602454378332713+0.9999102054675333i$ & & \\
& $-1$ & $-0.3602454378332713-0.9999102054675333i$ & & \\
\midrule
$0_3$ & - & $0$ & $-1.5$ & $0$ & $0$ \\
& - & $i$ & & \\
& - & $-i$ & & \\
\midrule
$1_3$ & $1$ & $0.3600822571547615$ & $-1.3702226735373758$ & $1$ & $0.0000456202502845$ \\
& $1$ & $0.3602454378332713+0.9999102054675333i$ & & \\
& $2$ & $0.3602454378332713-0.9999102054675333i$ & & \\
\midrule
$2_3$ & $2$ & $0.7780251893799998$ & $-1.0388285888128867$ & $2$ & $0.0005223047734948$ \\
& $1$ & $0.7884690768841303+1.0095363580274479i$ & & \\
& $2$ & $0.7884690768841303-1.0095363580274479i$ & & \\
\midrule
$3_3$ & $1$ & $1.3751320020322765$ & $-0.6339745962155614$ & $3$ & $0.0000663240402261$ \\
& $2$ & $1.2995316710963794+1.0960218400182442i$ & & \\
& $3$ & $1.2995316710963794-1.0960218400182442i$ & & \\
 \bottomrule
 \end{tabular}
 \captionsetup{width=.65\textwidth}
\caption{$N=6$, $M=3$, part II, 36 states}
\end{table}

\begin{table}[h!]
\scriptsize
\centering
\begin{tabular}{rrrrrr}
BT & B & $\lambda$& E & P & $|F^{zz}_{M4}|^2$ \\[0.3em]
\toprule
$-2.5_2$ & $-1.5$ & $-0.8036647561703933+0.5782572622173032i$ & $-3.3866956443879710$ & $2$ & $0.0170777128241272$ \\
& $-0.5$ & $-0.8036647561703933-0.5782572622173032i$ & & \\
$-1.5_2$ & $-1.5$ & $-0.2079830768876115+0.5062397808815254i$ & & \\
& $-0.5$ & $-0.2079830768876115-0.5062397808815254i$ & & \\
\midrule
$-2.5_2$ & $-1.5$ & $-0.8888704051389215+0.5754403251444594i$ & $-3.5687293044088437$ & $3$ & $0.1484547173477260$ \\
& $-0.5$ & $-0.8888704051389215-0.5754403251444594i$ & & \\
$-0.5_2$ & $-0.5$ & $0.0256601234818369+0.4967118046497673i$ & & \\
& $-0.5$ & $0.0256601234818369-0.4967118046497673i$ & & \\
\midrule
$-2.5_2$ & $-1.5$ & $-0.9621948944414796+0.5763832396210103i$ & $-3.0702431803750762$ & $4$ & $0.0434998349234136$ \\
& $-0.5$ & $-0.9621948944414796-0.5763832396210103i$ & & \\
$0.5_2$ & $-0.5$ & $0.2611251115948570+0.5045447969804691i$ & & \\
& $0.5$ & $0.2611251115948570-0.5045447969804691i$ & & \\
\midrule
$-2.5_2$ & $-1.5$ & $-1.0401751926378281+0.5795693023558029i$ & $-2.1746745323454712$ & $5$ & $0.0002049178477000$ \\
& $-0.5$ & $-1.0401751926378281-0.5795693023558029i$ & & \\
$1.5_2$ & $0.5$ & $0.6042632900130412+0.4766052282890622i$ & & \\
& $0.5$ & $0.6042632900130412-0.4766052282890622i$ & & \\
\midrule
$-2.5_2$ & $-1.5$ & $-1.1345148425061558+0.5848723992472916i$ & $-1.4032554973559433$ & $0$ & $0.0000000000000000$ \\
& $-0.5$ & $-1.1345148425061558-0.5848723992472916i$ & & \\
$2.5_2$ & $0.5$ & $1.1345148425061558+0.5848723992472916i$ & & \\
& $1.5$ & $1.1345148425061558-0.5848723992472916i$ & & \\
\midrule
$-1.5_2$ & $-0.5$ & $-0.3969559927993647+0.4607080317005712i$ & $-4.4478315142860692$ & $4$ & $0.9080091111358645$ \\
& $-0.5$ & $-0.3969559927993647-0.4607080317005712i$ & & \\
$-0.5_2$ & $-0.5$ & $-0.0561473304008010+0.4900443352694275i$ & & \\
& $-0.5$ & $-0.0561473304008010-0.4900443352694275i$ & & \\
\midrule
$-1.5_2$ & $-0.5$ & $-0.4642107980156561+0.4742493249967691i$ & $-4.0905762324398553$ & $5$ & $0.3270269459574597$ \\
& $-0.5$ & $-0.4642107980156561-0.4742493249967691i$ & & \\
$0.5_2$ & $-0.5$ & $0.1949955740300619+0.5060832884369909i$ & & \\
& $0.5$ & $0.1949955740300619-0.5060832884369909i$ & & \\
\midrule
$-1.5_2$ & $-0.5$ & $-0.5263290621701776+0.4768669715419894i$ & $-3.1015212750487114$ & $0$ & $0.0000000000000000$ \\
& $-0.5$ & $-0.5263290621701776-0.4768669715419894i$ & & \\
$1.5_2$ & $0.5$ & $0.5263290621701776+0.4768669715419894i$ & & \\
& $0.5$ & $0.5263290621701776-0.4768669715419894i$ & & \\
\midrule
$-1.5_2$ & $-0.5$ & $-0.6042632900130412+0.4766052282890622i$ & $-2.1746745323454712$ & $1$ & $0.0002049178477000$ \\
& $-0.5$ & $-0.6042632900130412-0.4766052282890622i$ & & \\
$2.5_2$ & $0.5$ & $1.0401751926378281+0.5795693023558029i$ & & \\
& $1.5$ & $1.0401751926378281-0.5795693023558029i$ & & \\
\midrule
$-0.5_2$ & $-0.5$ & $-0.1492571102680958+0.5101771962770252i$ & $-5.0477319188686440$ & $0$ & $-$(GS) \\
& $0.5$ & $-0.1492571102680958-0.5101771962770252i$ & & \\
$0.5_2$ & $0.5$ & $0.1492571102680958+0.5101771962770252i$ & & \\
& $0.5$ & $0.1492571102680958-0.5101771962770252i$ & & \\
\midrule
$-0.5_2$ & $-0.5$ & $-0.1949955740300619+0.5060832884369909i$ & $-4.0905762324398553$ & $1$ & $0.3270269459574597$ \\
& $0.5$ & $-0.1949955740300619-0.5060832884369909i$ & & \\
$1.5_2$ & $0.5$ & $0.4642107980156561+0.4742493249967691i$ & & \\
& $0.5$ & $0.4642107980156561-0.4742493249967691i$ & & \\
\midrule
$-0.5_2$ & $-0.5$ & $-0.2611251115948570+0.5045447969804691i$ & $-3.0702431803750762$ & $2$ & $0.0434998349234136$ \\
& $0.5$ & $-0.2611251115948570-0.5045447969804691i$ & & \\
$2.5_2$ & $0.5$ & $0.9621948944414796+0.5763832396210103i$ & & \\
& $1.5$ & $0.9621948944414796-0.5763832396210103i$ & & \\
\midrule
$0.5_2$ & $0.5$ & $0.0561473304008010+0.4900443352694275i$ & $-4.4478315142860692$ & $2$ & $0.9080091111358645$ \\
& $0.5$ & $0.0561473304008010-0.4900443352694275i$ & & \\
$1.5_2$ & $0.5$ & $0.3969559927993647+0.4607080317005712i$ & & \\
& $0.5$ & $0.3969559927993647-0.4607080317005712i$ & & \\
\midrule
$0.5_2$ & $0.5$ & $-0.0256601234818369+0.4967118046497673i$ & $-3.5687293044088437$ & $3$ & $0.1484547173477260$ \\
& $0.5$ & $-0.0256601234818369-0.4967118046497673i$ & & \\
$2.5_2$ & $0.5$ & $0.8888704051389215+0.5754403251444594i$ & & \\
& $1.5$ & $0.8888704051389215-0.5754403251444594i$ & & \\
\midrule
$1.5_2$ & $0.5$ & $0.2079830768876115+0.5062397808815254i$ & $-3.3866956443879710$ & $4$ & $0.0170777128241272$ \\
& $1.5$ & $0.2079830768876115-0.5062397808815254i$ & & \\
$2.5_2$ & $0.5$ & $0.8036647561703933+0.5782572622173032i$ & & \\
& $1.5$ & $0.8036647561703933-0.5782572622173032i$ & & \\
 \bottomrule
 \end{tabular}
 \captionsetup{width=.65\textwidth}
\caption{$N=6$, $M=4$, part I, 40 states}
\end{table}

\begin{table}[h!]
\scriptsize
\centering
\begin{tabular}{rrrrrr}
BT & B & $\lambda$& E & P & $|F^{zz}_{M4}|^2$\\[0.3em]
\toprule
$-1_1$ & $-1.5$ & $-0.4172174150276393$ & $-1.75$ & $0$ & $0.0000000000000000$ \\
$-2_3$ & $-0.5$ & $-0.8879962412278256$ & & \\
& $-2.5$ & $-0.8287589080915323+1.0547936386346191i$ & & \\
& $-1.5$ & $-0.8287589080915323-1.0547936386346191i$ & & \\
\midrule
$-1_1$ & $-0.5$ & $-0.7285638156760314$ & $-2.0709148404638211$ & $1$ & $0.0000265485047544$ \\
$-1_3$ & $-0.5$ & $-0.2843422676673447$ & & \\
& $-2.5$ & $-0.2842375171257300+0.9998124554859105i$ & & \\
& $-1.5$ & $-0.2842375171257300-0.9998124554859105i$ & & \\
\midrule
$-1_1$ & $-0.5$ & $-0.9139696370417966$ & $-2.0215473328758938$ & $2$ & $0.0006524833900225$ \\
$0_3$ & $0.5$ & $0.1453459697349351$ & & \\
& $0.5$ & $0.1453478261557062+0.9999999037165379i$ & & \\
& $1.5$ & $0.1453478261557062-0.9999999037165379i$ & & \\
\midrule
$-1_1$ & $-0.5$ & $-1.0436453670563748$ & $-1.6812706955911563$ & $3$ & $0.0005197183559271$ \\
$1_3$ & $1.5$ & $0.5734798372122437$ & & \\
& $0.5$ & $0.5756398390705960+1.0036242925593361i$ & & \\
& $1.5$ & $0.5756398390705960-1.0036242925593361i$ & & \\
\midrule
$-1_1$ & $-0.5$ & $-1.1590824419781914$ & $-1.1982838595313864$ & $4$ & $0.0001310462344822$ \\
$2_3$ & $0.5$ & $1.1352502921259584$ & & \\
& $1.5$ & $1.0779692771616892+1.0547382592318158i$ & & \\
& $2.5$ & $1.0779692771616892-1.0547382592318158i$ & & \\
\midrule
$0_1$ & $-0.5$ & $0.2535457639517348$ & $-1.7714707163348150$ & $1$ & $0.0006903212695128$ \\
$-2_3$ & $-0.5$ & $-1.0316371808126689$ & & \\
& $-2.5$ & $-0.9777074330195156+1.0481020833265213i$ & & \\
& $-1.5$ & $-0.9777074330195156-1.0481020833265213i$ & & \\
\midrule
$0_1$ & $-0.5$ & $0.1403957698390096$ & $-2.2689018975264362$ & $2$ & $0.0272280200506845$ \\
$-1_3$ & $-1.5$ & $-0.4603172854240659$ & & \\
& $-1.5$ & $-0.4615442327573274+1.0023724999658045i$ & & \\
& $-0.5$ & $-0.4615442327573274-1.0023724999658045i$ & & \\
\midrule
$0_1$ & $-$ & $0$ & $-2.5$ & $0$ & $0$ \\
$0_3$ & $-$ & $0$ & & \\
& $-$ & $i$ & & \\
& $-$ & $-i$ & & \\
\midrule
$0_1$ & $0.5$ & $-0.1403957698390096$ & $-2.2689018975264362$ & $4$ & $0.0272280200506845$ \\
$1_3$ & $1.5$ & $0.4603172854240659$ & & \\
& $0.5$ & $0.4615442327573274+1.0023724999658045i$ & & \\
& $1.5$ & $0.4615442327573274-1.0023724999658045i$ & & \\
\midrule
$0_1$ & $0.5$ & $-0.2535457639517348$ & $-1.7714707163348150$ & $5$ & $0.0006903212695128$ \\
$2_3$ & $0.5$ & $1.0316371808126689$ & & \\
& $1.5$ & $0.9777074330195156+1.0481020833265213i$ & & \\
& $2.5$ & $0.9777074330195156-1.0481020833265213i$ & & \\
\midrule
$1_1$ & $0.5$ & $1.1590824419781914$ & $-1.1982838595313864$ & $2$ & $0.0001310462344822$ \\
$-2_3$ & $-0.5$ & $-1.1352502921259584$ & & \\
& $-2.5$ & $-1.0779692771616892+1.0547382592318158i$ & & \\
& $-1.5$ & $-1.0779692771616892-1.0547382592318158i$ & & \\
\midrule
$1_1$ & $0.5$ & $1.0436453670563748$ & $-1.6812706955911563$ & $3$ & $0.0005197183559271$ \\
$-1_3$ & $-1.5$ & $-0.5734798372122437$ & & \\
& $-1.5$ & $-0.5756398390705960+1.0036242925593361i$ & & \\
& $-0.5$ & $-0.5756398390705960-1.0036242925593361i$ & & \\
\midrule
$1_1$ & $0.5$ & $0.9139696370417966$ & $-2.0215473328758938$ & $4$ & $0.0006524833900225$ \\
$0_3$ & $-0.5$ & $-0.1453459697349351$ & & \\
& $-1.5$ & $-0.1453478261557062+0.9999999037165379i$ & & \\
& $-0.5$ & $-0.1453478261557062-0.9999999037165379i$ & & \\
\midrule
$1_1$ & $0.5$ & $0.7285638156760314$ & $-2.0709148404638211$ & $5$ & $0.0000265485047544$ \\
$1_3$ & $0.5$ & $0.2843422676673447$ & & \\
& $1.5$ & $0.2842375171257300+0.9998124554859105i$ & & \\
& $2.5$ & $0.2842375171257300-0.9998124554859105i$ & & \\
\midrule
$1_1$ & $1.5$ & $0.4172174150276393$ & $-1.75$ & $0$ & $0.0000000000000000$ \\
$2_3$ & $0.5$ & $0.8879962412278256$ & & \\
& $1.5$ & $0.8287589080915323+1.0547936386346191i$ & & \\
& $2.5$ & $0.8287589080915323-1.0547936386346191i$ & & \\
 \bottomrule
 \end{tabular}
 \captionsetup{width=.65\textwidth}
\caption{$N=6$, $M=4$, part II, 40 states}
\end{table}
\begin{table}[h!]
\scriptsize
\centering
\begin{tabular}{rrrrrr}
BT & B & $\lambda$& E & P & $|F^{zz}_{M4}|^2$ \\[0.3em]
\toprule
$0.5_1$ & $0.5$ & $0.9786507997042435$ & $-2.3672139253321358$ & $4$ & $0.0003384867317601$ \\
$-0.5_1$ & $-0.5$ & $-0.1483156944498343$ & & \\
$-2_2$ & $-1.5$ & $-0.9254685761284599+0.5798860423518019i$ & & \\
& $-0.5$ & $-0.9254685761284599-0.5798860423518019i$ & & \\
\midrule
$0.5_1$ & $0.5$ & $0.9004363395453612$ & $-3.4137102060124348$ & $5$ & $0.0005159464982604$ \\
$-0.5_1$ & $-0.5$ & $-0.6398190009845127$ & & \\
$-1_2$ & $-0.5$ & $-0.2803753366392453+0.4883202503674975i$ & & \\
& $-0.5$ & $-0.2803753366392453-0.4883202503674975i$ & & \\
\midrule
$0.5_1$ & $0.5$ & $0.8195774644159196$ & $-3.8778480615673433$ & $0$ & $0.0000000000000000$ \\
$-0.5_1$ & $-0.5$ & $-0.8195774644159196$ & & \\
$0_2$ & $-0.5$ & $0.5041224339899170i$ & & \\
& $0.5$ & $-0.5041224339899170i$ & & \\
\midrule
$0.5_1$ & $0.5$ & $0.6398190009845127$ & $-3.4137102060124348$ & $1$ & $0.0005159464982604$ \\
$-0.5_1$ & $-0.5$ & $-0.9004363395453612$ & & \\
$1_2$ & $0.5$ & $0.2803753366392453+0.4883202503674975i$ & & \\
& $0.5$ & $0.2803753366392453-0.4883202503674975i$ & & \\
\midrule
$0.5_1$ & $0.5$ & $0.1483156944498343$ & $-2.3672139253321358$ & $2$ & $0.0003384867317601$ \\
$-0.5_1$ & $-0.5$ & $-0.9786507997042435$ & & \\
$2_2$ & $0.5$ & $0.9254685761284599+0.5798860423518019i$ & & \\
& $1.5$ & $0.9254685761284599-0.5798860423518019i$ & & \\
\midrule
$-2_4$ & $-1.5$ & $-1.1036662393572179+0.5078701404301931i$ & $-0.7392826456850315$ & $4$ & $0.0000584860163966$ \\
& $-0.5$ & $-1.1036662393572179-0.5078701404301931i$ & & \\
& $-3.5$ & $-1.0360550682203347+1.4880380646500060i$ & & \\
& $-2.5$ & $-1.0360550682203347-1.4880380646500060i$ & & \\
\midrule
$-1_4$ & $-0.5$ & $-0.5079588269222967+0.4999303471007927i$ & $-0.9786534724036026$ & $5$ & $0.0000884032763060$ \\
& $-0.5$ & $-0.5079588269222967-0.4999303471007927i$ & & \\
& $-3.5$ & $-0.5056159447610890+1.4948984128570630i$ & & \\
& $-2.5$ & $-0.5056159447610890-1.4948984128570630i$ & & \\
\midrule
$0_4$ & $-0.5$ & $0.5000015897353024i$ & $-1.0696432471593580$ & $0$ & $0.0000000000000000$ \\
& $0.5$ & $-0.5000015897353024i$ & & \\
& $1.5$ & $1.5007749182844950i$ & & \\
& $-1.5$ & $-1.5007749182844950i$ & & \\
\midrule
$1_4$ & $0.5$ & $0.5079588269222967+0.4999303471007927i$ & $-0.9786534724036026$ & $1$ & $0.0000884032763060$ \\
& $0.5$ & $0.5079588269222967-0.4999303471007927i$ & & \\
& $2.5$ & $0.5056159447610890+1.4948984128570630i$ & & \\
& $3.5$ & $0.5056159447610890-1.4948984128570630i$ & & \\
\midrule
$2_4$ & $0.5$ & $1.1036662393572179+0.5078701404301931i$ & $-0.7392826456850315$ & $2$ & $0.0000584860163966$ \\
& $1.5$ & $1.1036662393572179-0.5078701404301931i$ & & \\
& $2.5$ & $1.0360550682203347+1.4880380646500060i$ & & \\
& $3.5$ & $1.0360550682203347-1.4880380646500060i$ & & \\
 \bottomrule
 \end{tabular}
 \captionsetup{width=.65\textwidth}
\caption{$N=6$, $M=4$, part III 40 states}
\end{table}
\newpage
\begin{table}[h!]
\scriptsize
\centering
\begin{tabular}{rrrrrr}
BT & B & $\lambda$& E & P & $|F^{-+}_{M6}|^2$  \\[0.3em]
\toprule
$0_1$ & $0$ & $-0.6155824896741097$ & $-4.6674217796794355$ & $5$ & $0.4621517596557141$ \\
$0.5_2$ & $0.0$ & $0.1042754099415518+0.5098278117754908i$ & & \\
& $1.0$ & $0.1042754099415518-0.5098278117754908i$ & & \\
$1.5_2$ & $0.0$ & $0.6113014527883884+0.5600502322371753i$ & & \\
& $1.0$ & $0.6113014527883884-0.5600502322371753i$ & & \\
\midrule
$0_1$ & $0$ & $-0.4124492474611392$ & $-4.4474281868168146$ & $4$ & $1.2305819964533264$ \\
$-0.5_2$ & $0.0$ & $-0.1906025411394349+0.4848058776348274i$ & & \\
& $0.0$ & $-0.1906025411394349-0.4848058776348274i$ & & \\
$1.5_2$ & $0.0$ & $0.6742292028878124+0.5553405201370433i$ & & \\
& $1.0$ & $0.6742292028878124-0.5553405201370433i$ & & \\
\midrule
$0_1$ & $0$ & $0$ & $-5.6629176259105471$ & $3$ & $4.4068019602365224$ \\
$-0.5_2$ & $0.0$ & $-0.1777470423567636+0.4529111371009282i$ & & \\
& $0.0$ & $-0.1777470423567636-0.4529111371009282i$ & & \\
$0.5_2$ & $0.0$ & $0.1777470423567636+0.4529111371009282i$ & & \\
& $0.0$ & $0.1777470423567636-0.4529111371009282i$ & & \\
\midrule
$0_1$ & $0$ & $0$ & $-3.2555419656828377$ & $3$ & $0.1608850207944900$ \\
$-1.5_2$ & $-1.0$ & $-0.7344347264939947+0.5567522611426832i$ & & \\
& $0.0$ & $-0.7344347264939947-0.5567522611426832i$ & & \\
$1.5_2$ & $0.0$ & $0.7344347264939947+0.5567522611426832i$ & & \\
& $1.0$ & $0.7344347264939947-0.5567522611426832i$ & & \\
\midrule
$0_1$ & $0$ & $0.4124492474611392$ & $-4.4474281868168146$ & $2$ & $1.2305819964533264$ \\
$-1.5_2$ & $-1.0$ & $-0.6742292028878124+0.5553405201370433i$ & & \\
& $0.0$ & $-0.6742292028878124-0.5553405201370433i$ & & \\
$0.5_2$ & $0.0$ & $0.1906025411394349+0.4848058776348274i$ & & \\
& $0.0$ & $0.1906025411394349-0.4848058776348274i$ & & \\
\midrule
$0_1$ & $0$ & $0.6155824896741097$ & $-4.6674217796794355$ & $1$ & $0.4621517596557141$ \\
$-1.5_2$ & $-1.0$ & $-0.6113014527883884+0.5600502322371753i$ & & \\
& $0.0$ & $-0.6113014527883884-0.5600502322371753i$ & & \\
$-0.5_2$ & $-1.0$ & $-0.1042754099415518+0.5098278117754908i$ & & \\
& $0.0$ & $-0.1042754099415518-0.5098278117754908i$ & & \\
\midrule
$-0.5_1$ & $0$ & $0.7760928156236603$ & $-2.6535245095478798$ & $5$ & $0.0007866560631482$ \\
$0.5_1$ & $-1$ & $-0.3361920951945765$ & & \\
$-1_3$ & $0$ & $-0.6116744597578116$ & & \\
& $-2$ & $-0.5774906107067409+1.0182692877505771i$ & & \\
& $-1$ & $-0.5774906107067409-1.0182692877505771i$ & & \\
\midrule
$-0.5_1$ & $-$ & $0.6151741139620555$ & $-2.9509163797718942$ & $0$ & $0$ \\
$0.5_1$ & $-$ & $-0.6151741139620555$ & & \\
$0_3$ & $-$ & $0$ & & \\
& $-$ & $i$ & & \\
& $-$ & $-i$ & & \\
\midrule
$-0.5_1$ & $1$ & $0.3361920951945765$ & $-2.6535245095478798$ & $1$ & $0.0007866560631482$ \\
$0.5_1$ & $0$ & $-0.7760928156236603$ & & \\
$1_3$ & $0$ & $0.6116744597578116$ & & \\
& $1$ & $0.5774906107067409+1.0182692877505771i$ & & \\
& $2$ & $0.5774906107067409-1.0182692877505771i$ & & \\
\midrule
$0_2$ & $-1$ & $0.1475046631495792+0.5044326797467416i$ & $-3.6337545340012151$ & $5$ & $0.0029471263324903$ \\
& $0$ & $0.1475046631495792-0.5044326797467416i$ & & \\
$-1_3$ & $0$ & $-0.6640449056199454$ & & \\
& $-2$ & $-0.6419946016559088+1.0097957534813132i$ & & \\
& $-1$ & $-0.6419946016559088-1.0097957534813132i$ & & \\
\midrule
$0_2$ & $-$ & $0.50638053617925371i$ & $-4.1896949288340794$ & $0$ & $0$ \\
& $-$ & $-0.50638053617925371i$ & & \\
$0_3$ & $-$ & $0$ & & \\
& $-$ & $i$ & & \\
& $-$ & $-i$ & & \\
\midrule
$0_2$ & $0$ & $-0.1475046631495792+0.5044326797467416i$ & $-3.6337545340012151$ & $1$ & $0.0029471263324903$ \\
& $1$ & $-0.1475046631495792-0.5044326797467416i$ & & \\
$1_3$ & $0$ & $0.6640449056199454$ & & \\
& $1$ & $0.6419946016559088+1.0097957534813132i$ & & \\
& $2$ & $0.6419946016559088-1.0097957534813132i$ & & \\
 \bottomrule
 \end{tabular}
 \captionsetup{width=.65\textwidth}
\caption{$N=6$, $M=5$, part I, 36 states}
\end{table}
\newpage
\begin{table}[h!]
\scriptsize
\centering
\begin{tabular}{rrrrrr}
BT & B & $\lambda$& E & P & $|F^{-+}_{M6}|^2$ \\[0.3em]
\toprule
$-2_2$ & $-1$ & $-0.5906005172425210+0.5405390439370296i$ & $-2.8430703308172536$ & $3$ & $0.0033239138830987$ \\
& $0$ & $-0.5906005172425210-0.5405390439370296i$ & & \\
$-1_3$ & $0$ & $-0.1985520765620395$ & & \\
& $-3$ & $-0.1985109822637059+1.0000258094976104i$ & & \\
& $-2$ & $-0.1985109822637059-1.0000258094976104i$ & & \\
\midrule
$-2_2$ & $-1$ & $-0.7931934831195145+0.5517636692053256i$ & $-2.4367936379920076$ & $4$ & $0.0016649380829351$ \\
& $0$ & $-0.7931934831195145-0.5517636692053256i$ & & \\
$0_3$ & $1$ & $0.3288644919332310$ & & \\
& $0$ & $0.3288964591888249+1.0003559732705106i$ & & \\
& $1$ & $0.3288964591888249-1.0003559732705106i$ & & \\
\midrule
$-2_2$ & $-1$ & $-0.9251983995938861+0.5599681537480929i$ & $-1.8519879291949645$ & $5$ & $0.0000925679194070$ \\
& $0$ & $-0.9251983995938861-0.5599681537480929i$ & & \\
$1_3$ & $0$ & $0.8760411873717446$ & & \\
& $1$ & $0.8418101696958006+1.0205436836407630i$ & & \\
& $2$ & $0.8418101696958006-1.0205436836407630i$ & & \\
\midrule
$-1_2$ & $-1$ & $-0.1091788921019517+0.4936601494699574i$ & $-3.7730004341516686$ & $4$ & $0.0109185293344029$ \\
& $-1$ & $-0.1091788921019517-0.4936601494699574i$ & & \\
$-1_3$ & $0$ & $-0.5391270914300524$ & & \\
& $-2$ & $-0.5252910935702397+1.0059280271340690i$ & & \\
& $-1$ & $-0.5252910935702397-1.0059280271340690i$ & & \\
\midrule
$-1_2$ & $0$ & $-0.3370523045990830+0.4854883632265467i$ & $-3.4608271845507004$ & $5$ & $0.0035697160123478$ \\
& $0$ & $-0.3370523045990830-0.4854883632265467i$ & & \\
$0_3$ & $-1$ & $0.1878240259216440$ & & \\
& $1$ & $0.1878233539215883+1.0000319224320813i$ & & \\
& $2$ & $0.1878233539215883-1.0000319224320813i$ & & \\
\midrule
$-1_2$ & $0$ & $-0.4550876892796446+0.4847583844937834i$ & $-2.75$ & $0$ & $0.0000000000000000$ \\
& $0$ & $-0.4550876892796446-0.4847583844937834i$ & & \\
$1_3$ & $0$ & $0.7629784368524934$ & & \\
& $1$ & $0.7350362986195455+1.0141304262451267i$ & & \\
& $2$ & $0.7350362986195455-1.0141304262451267i$ & & \\
\midrule
$1_2$ & $0$ & $0.4550876892796446+0.4847583844937834i$ & $-2.75$ & $0$ & $0.0000000000000000$ \\
& $0$ & $0.4550876892796446-0.4847583844937834i$ & & \\
$-1_3$ & $0$ & $-0.7629784368524934$ & & \\
& $-2$ & $-0.7350362986195455+1.0141304262451267i$ & & \\
& $-1$ & $-0.7350362986195455-1.0141304262451267i$ & & \\
\midrule
$1_2$ & $0$ & $0.3370523045990830+0.4854883632265467i$ & $-3.4608271845507004$ & $1$ & $0.0035697160123478$ \\
& $0$ & $0.3370523045990830-0.4854883632265467i$ & & \\
$0_3$ & $1$ & $-0.1878240259216440$ & & \\
& $-2$ & $-0.1878233539215883+1.0000319224320813i$ & & \\
& $-1$ & $-0.1878233539215883-1.0000319224320813i$ & & \\
\midrule
$1_2$ & $1$ & $0.1091788921019517+0.4936601494699574i$ & $-3.7730004341516686$ & $2$ & $0.0109185293344029$ \\
& $1$ & $0.1091788921019517-0.4936601494699574i$ & & \\
$1_3$ & $0$ & $0.5391270914300524$ & & \\
& $1$ & $0.5252910935702397+1.0059280271340690i$ & & \\
& $2$ & $0.5252910935702397-1.0059280271340690i$ & & \\
\midrule
$2_2$ & $0$ & $0.9251983995938861+0.5599681537480929i$ & $-1.8519879291949645$ & $1$ & $0.0000925679194070$ \\
& $1$ & $0.9251983995938861-0.5599681537480929i$ & & \\
$-1_3$ & $0$ & $-0.8760411873717446$ & & \\
& $-2$ & $-0.8418101696958006+1.0205436836407630i$ & & \\
& $-1$ & $-0.8418101696958006-1.0205436836407630i$ & & \\
\midrule
$2_2$ & $0$ & $0.7931934831195145+0.5517636692053256i$ & $-2.4367936379920076$ & $2$ & $0.0016649380829351$ \\
& $1$ & $0.7931934831195145-0.5517636692053256i$ & & \\
$0_3$ & $-1$ & $-0.3288644919332310$ & & \\
& $-1$ & $-0.3288964591888249+1.0003559732705106i$ & & \\
& $0$ & $-0.3288964591888249-1.0003559732705106i$ & & \\
\midrule
$2_2$ & $0$ & $0.5906005172425210+0.5405390439370296i$ & $-2.8430703308172536$ & $3$ & $0.0033239138830987$ \\
& $1$ & $0.5906005172425210-0.5405390439370296i$ & & \\
$1_3$ & $0$ & $0.1985520765620395$ & & \\
& $2$ & $0.1985109822637059+1.0000258094976104i$ & & \\
& $3$ & $0.1985109822637059-1.0000258094976104i$ & & \\
 \bottomrule
 \end{tabular}
 \captionsetup{width=.65\textwidth}
\caption{$N=6$, $M=5$, part II, 36 states}
\end{table}
\begin{table}[h!]
\scriptsize
\centering
\begin{tabular}{rrrrrrr}
BT & B & $\lambda$& E & P & $|F^{-+}_{M6}|^2$ \\[0.3em]
\toprule
$-1_1$ & $0$ & $-0.5927610624475779$ & $-1.6988236863193494$ & $1$ & $0.0000075322750541$ \\
$-1_4$ & $-1$ & $-0.5012018625258346+0.5008757902114886i$ & & \\
& $0$ & $-0.5012018625258346-0.5008757902114886i$ & & \\
& $-4$ & $-0.4871697834582320+1.4812325187084924i$ & & \\
& $-3$ & $-0.4871697834582320-1.4812325187084924i$ & & \\
\midrule
$-1_1$ & $0$ & $-0.8069757503682298$ & $-1.6731795677341720$ & $2$ & $0.0000531817392344$ \\
$0_4$ & $1$ & $0.1470891108376228+0.4999895733158748i$ & & \\
& $1$ & $0.1470891108376228-0.4999895733158748i$ & & \\
& $1$ & $0.1485797086290681+1.5019573893925587i$ & & \\
& $2$ & $0.1485797086290681-1.5019573893925587i$ & & \\
\midrule
$-1_1$ & $0$ & $-0.9638386461100639$ & $-1.4069296691827464$ & $3$ & $0.0000163035209596$ \\
$1_4$ & $0$ & $0.7853472319703160+0.5019690643465596i$ & & \\
& $1$ & $0.7853472319703160-0.5019690643465596i$ & & \\
& $2$ & $0.7681445723963226+1.4697036613175872i$ & & \\
& $3$ & $0.7681445723963226-1.4697036613175872i$ & & \\
\midrule
$0_1$ & $-1$ & $0.1290830017729629$ & $-1.9195981733053373$ & $2$ & $0.0000155680560221$ \\
$-1_4$ & $-1$ & $-0.6626985823965715+0.5015934941027946i$ & & \\
& $0$ & $-0.6626985823965715-0.5015934941027946i$ & & \\
& $-3$ & $-0.6535328359981892+1.4689553023145136i$ & & \\
& $-2$ & $-0.6535328359981892-1.4689553023145136i$ & & \\
\midrule
$0_1$ & $0$ & $0$ & $-2.0815404084066153$ & $3$ & $0.0000533546544947$ \\
$0_4$ & $0$ & $0.4999757102424680i$ & & \\
& $0$ & $-0.4999757102424680i$ & & \\
& $1$ & $1.5039274881011447i$ & & \\
& $-1$ & $-1.5039274881011447i$ & & \\
\midrule
$0_1$ & $1$ & $-0.1290830017729629$ & $-1.9195981733053373$ & $4$ & $0.0000155680560221$ \\
$1_4$ & $0$ & $0.6626985823965715+0.5015934941027946i$ & & \\
& $1$ & $0.6626985823965715-0.5015934941027946i$ & & \\
& $2$ & $0.6535328359981892+1.4689553023145136i$ & & \\
& $3$ & $0.6535328359981892-1.4689553023145136i$ & & \\
\midrule
$1_1$ & $0$ & $0.9638386461100639$ & $-1.4069296691827464$ & $3$ & $0.0000163035209596$ \\
$-1_4$ & $-1$ & $-0.7853472319703160+0.5019690643465596i$ & & \\
& $0$ & $-0.7853472319703160-0.5019690643465596i$ & & \\
& $-3$ & $-0.7681445723963226+1.4697036613175872i$ & & \\
& $-2$ & $-0.7681445723963226-1.4697036613175872i$ & & \\
\midrule
$1_1$ & $0$ & $0.8069757503682298$ & $-1.6731795677341720$ & $4$ & $0.0000531817392344$ \\
$0_4$ & $-1$ & $-0.1470891108376228+0.4999895733158748i$ & & \\
& $-1$ & $-0.1470891108376228-0.4999895733158748i$ & & \\
& $-2$ & $-0.1485797086290681+1.5019573893925587i$ & & \\
& $-1$ & $-0.1485797086290681-1.5019573893925587i$ & & \\
\midrule
$1_1$ & $0$ & $0.5927610624475779$ & $-1.6988236863193494$ & $5$ & $0.0000075322750541$ \\
$1_4$ & $0$ & $0.5012018625258346+0.5008757902114886i$ & & \\
& $1$ & $0.5012018625258346-0.5008757902114886i$ & & \\
& $3$ & $0.4871697834582320+1.4812325187084924i$ & & \\
& $4$ & $0.4871697834582320-1.4812325187084924i$ & & \\
\midrule
$-1_5$ & $-2$ & $-0.7023744289880658$ & $-0.7836603767064553$ & $5$ & $0.0000000428291057$ \\
& $-3$ & $-0.7020556998129408+0.9994152751081230i$ & & \\
& $-2$ & $-0.7020556998129408-0.9994152751081230i$ & & \\
& $-2$ & $-0.7477585408541322+1.9339783678268855i$ & & \\
& $-1$ & $-0.7477585408541322-1.9339783678268855i$ & & \\
\midrule
$0_5$ & $-$ & $0$ & $-0.8593886913940265$ & $0$ & $0$ \\
& $-$ & $i$ & & \\
& $-$ & $-i$ & & \\
& $-$ & $2.0302753367629257i$ & & \\
& $-$ & $-2.0302753367629257i$ & & \\
\midrule
$1_5$ & $2$ & $0.7023744289880658$ & $-0.7836603767064553$ & $1$ & $0.0000000428291057$ \\
& $2$ & $0.7020556998129408+0.9994152751081230i$ & & \\
& $3$ & $0.7020556998129408-0.9994152751081230i$ & & \\
& $1$ & $0.7477585408541322+1.9339783678268855i$ & & \\
& $2$ & $0.7477585408541322-1.9339783678268855i$ & & \\
 \bottomrule
 \end{tabular}
 \captionsetup{width=.65\textwidth}
\caption{$N=6$, $M=5$, part III, 36 states}
\end{table}
\end{document}